\documentclass[%
 reprint,
 amsmath,amssymb,
 aps,
]{revtex4-2}
\usepackage[colorlinks,linkcolor=blue,urlcolor=blue,citecolor=blue]{hyperref}
\usepackage{graphicx}% Include figure files
\usepackage{dcolumn}% Align table columns on decimal point
\usepackage{bm}% bold math
\usepackage{subfigure}
\begin{document}

\preprint{APS/123-QED}

\title{Theoretical and experimental studies of energy modulation to demodulation \\in seeded free-electron lasers}

\author{Hanxiang Yang\textsuperscript{1}}
\author{Nanshun Huang\textsuperscript{1}}
\author{Zipeng Liu\textsuperscript{2}}
\author{Zhangfeng Gao\textsuperscript{1}}
\author{Shengbin Ye\textsuperscript{2}}
\author{Wencai Cheng\textsuperscript{1}}
\author{Shudong Zhou\textsuperscript{1}}
\author{Jinya Chen\textsuperscript{1}}
\author{Cheng Yu\textsuperscript{1}}
\author{Wei Zhang\textsuperscript{1}}
\author{Tao Liu\textsuperscript{1}}
% \author{Wei Zhang\textsuperscript{1}}
\author{Haixiao Deng\textsuperscript{1,}}%
\email{denghx@sari.ac.cn}
\affiliation{\textsuperscript{1}Shanghai Advanced Research Institute, Chinese Academy of Sciences, Shanghai 201210, China\\
	\textsuperscript{2}Zhangjiang Laboratory, Shanghai 201210, China\\
\\
}%

\date{\today}% It is always \today, today,

\begin{abstract}
Laser manipulation plays a critical role in precisely tailoring relativistic electron beams through energy modulation, enabling the generation of coherent, intense, and ultrashort radiation in accelerator-based light sources such as synchrotron radiation facilities and free-electron lasers (FELs). However, laser-induced energy modulation inevitably degrades electron beam quality by increasing the energy spread, thereby limiting high-repetition-rate operation. Here, we investigate energy modulation and demodulation in a seeded FEL using two modulators separated by a tunable phase shifter. Analytical analysis and three-dimensional simulations show that a $\pi$ phase delay can nearly reverse the laser-beam interaction and substantially suppress the residual modulation. Diagnostics based on coherent undulator radiation and time-resolved measurements are established to characterize weak residual modulation, and a dedicated demodulation undulator is designed for controlled studies. Preliminary experiments performed at the Shanghai soft X-ray FEL facility using the existing seeding beamline demonstrate laser-induced energy-modulation suppression. Together with the analytical and numerical studies, these results establish a practical framework for investigating the transition from energy modulation to demodulation in seeded FELs, with potential applications in high-repetition-rate, fully coherent X-ray sources with improved preservation of electron beam quality.
\end{abstract}
\maketitle
%\tableofcontents
\section{\label{sec:1}Introduction}
Relativistic electron beams generated by particle accelerators underpin large-scale light sources, such as linac-based free-electron lasers (FELs) and storage ring-based synchrotron radiation facilities (SRs), enabling breakthroughs across materials science, biology, chemistry, and physics \citep{zhao2010,Pellegrini2016}. SR sources offer broad wavelength coverage, multi-user capability, and high stability, while FELs deliver high peak brightness, ultrashort pulses, and excellent longitudinal coherence \citep{Hastings2019,Huang2021}. The two technologies are advancing toward convergence, with the shared objective of realizing fully coherent X-ray sources at high repetition rates. Photon beam performance improvements in both FELs and SRs have been driven largely by laser-based manipulation of relativistic electron beams, which enables precise tailoring of the electron beam phase space and thus control over the emitted radiation \citep{Hemsing2014}.

In high-gain FELs, the self-amplified spontaneous emission (SASE) scheme is widely adopted but suffers from limited temporal coherence due to shot noise \citep{Kondratenko1980,BONIFACIO1984251,Bonifacio1994,Emma2010,Ishikawa2012,Kang2017,Decking2020,Prat2020}. Seeded FELs overcome this limit by using external lasers to imprint well-defined energy modulation and induce microbunching in the electron beam, thereby enabling the generation of fully coherent and spectrally pure extreme ultraviolet (EUV) and X-ray pulses \citep{feng2018review}. The coherent harmonic generation (CHG) scheme \citep{Yu1991} extends coherence into the ultraviolet regime. A more widely implemented approach, the high-gain harmonic generation (HGHG) scheme, enhances the harmonic bunching factor by placing a dispersive section after the modulator undulator \citep{Yu2000,Yu2003,Allaria2012}. Nevertheless, its performance is limited by the large laser-induced energy spread, which restricts efficient harmonic up-conversion to shorter wavelengths. To reach the soft X-ray regime, the cascaded HGHG scheme based on the “fresh bunch” technique was proposed and experimentally demonstrated, which poses significant challenges for extending operation below 4 nm \citep{Yu1997,Allaria2013,Liu2013}. Notably, through the laser–beam interaction in an undulator, the laser heater can suppress microbunching instability by increasing the beam’s uncorrelated energy spread, which is now essential in modern XFELs \citep{SALDIN2004355,Huang2004} and beneficial for externally seeded FELs \citep{Spampinati2014}. 

Several variants of the HGHG scheme, such as two-stage and multi-stage energy modulation configurations, have been investigated to further mitigate energy-spread growth while extending the output FEL wavelength. In these schemes, the electron beam interacts with intense seed lasers in two modulators separated by a phase shifter \citep{mcneil2005,Allaria2007,Jia2008}, typically tuned to a $\pi$ phase delay to enable partial demodulation between stages. In contrast, the echo-enabled harmonic generation (EEHG) scheme \citep{Stupakov2009,Xiang2009,Xiang2010X} employs two modulators and two dispersive sections to manipulate the electron beam’s longitudinal phase space in multiple dimensions, thereby achieving high harmonic generation with only modest energy-spread growth. EEHG and its cascaded implementations have demonstrated the capability to produce nearly Fourier-transform-limited soft X-ray FEL pulses \citep{Zhao2012,Hemsing2016,RebernikRibic2019,Feng2019,Penn2014,Feng2010,Feng22,Yang2022}. Moreover, the phase-merging enhanced harmonic generation scheme has been proposed to achieve remarkable harmonic up-conversion efficiency and reduced energy spread, benefiting from the transverse–longitudinal coupling of the electron beam phase space \citep{Deng2013,Feng_2014}. Additionally, the cascaded modulator–chicane module layout enables enhancement of the high-harmonic bunching factor in seeded FELs and has been applied in inverse FEL acceleration experiments \citep{xiang2011,Hemsing2013,SUDAR2017,Sudar2018}.

To improve the temporal coherence of SR sources, a series of laser-based electron-beam manipulation schemes has been proposed. Femtosecond laser modulation can generate ultrashort X-ray pulses, as demonstrated in laser-slicing experiments, though typically at the cost of increased energy spread \citep{Zolents1996,Schoenlein2000,Khan2006,Beaud2007}. Notably, advances in seeded FELs have inspired analogous storage-ring-based schemes for generating coherent EUV and X-ray pulses, including the CHG-based schemes \citep{LABAT2008,Ninno2008}, steady-state microbunching \citep{Ratner2010,deng2021experimental}, EEHG-based schemes \citep{Khan2013,Liu2018,Yang:yi5139}, and the angular-dispersion-induced microbunching scheme \citep{xiang2010,Feng2017,Li2020}. However, the laser-induced energy modulation process inevitably increases the energy spread and simultaneously enhances the vertical emittance due to the presence of nonzero vertical dispersion. A common approach to mitigate this effect splits the laser into two pulses with a $\pi$ phase difference, generating opposite modulations \citep{ratner2011reversible,tang2018overview}. Several demodulation schemes have also been proposed to restore beam quality, thereby sustaining stable multi-turn operation and high average power \citep{Jiang2022,Li2023,Liu2025}. Nevertheless, achieving effective modulation cancellation remains crucial for realizing high-repetition-rate, fully coherent SR-based light sources.

FELs and SRs are steadily progressing toward convergence. Laser-driven electron-beam energy modulation has emerged as a transformative technique for achieving the shared goal of fully coherent X-ray generation at high repetition rates. However, the energy-spread growth remains an unavoidable challenge, particularly at MHz-level repetition rates, where the achievable modulation amplitude in seeded FELs is limited by both the available seed-laser power and the intrinsic energy spread of the electron beam \citep{Ackermann2020,Yan2021,Yang2023,YANG2026705,qi2025}. Demodulation schemes---implemented by introducing a $\pi$ phase delay via a phase shifter---offer a promising approach to suppress residual energy modulation. Their practical realization, however, requires a quantitative understanding of the demodulation process, robust diagnostics for weak residual modulation, and experimental evidence under realistic seeded-FEL operating conditions.

% In this paper, we investigate energy modulation and demodulation in seeded FELs using a demodulation undulator configuration incorporating a phase shifter. Analytical estimates and three-dimensional simulations are performed using representative parameters of the Shanghai soft X-ray FEL facility (SXFEL) \citep{Liu2021}. The simulations quantify the achievable suppression of residual energy modulation and identify the diagnostic requirements for resolving weak modulation. Complementary measurement methods based on coherent undulator radiation and dispersion scan are analyzed. A dedicated demodulation undulator is also designed as a future platform for controlled demodulation experiments. Finally, measurements using the existing SXFEL seeding beamline are presented to show experimental signatures of energy modulation suppression.
In this paper, we investigate energy modulation and demodulation in seeded FELs using a demodulation undulator configuration incorporating a phase shifter. Unlike previous demodulator-based seeded FEL schemes \citep{mcneil2005,Allaria2007,Jia2008} aimed at preserving high-harmonic bunching for FEL amplification, this work focuses on the modulation-to-demodulation process itself and on the diagnosis of weak residual modulation. Analytical estimates and three-dimensional simulations are performed using representative parameters of the Shanghai soft X-ray FEL facility (SXFEL) \citep{Liu2021} to quantify the achievable suppression and identify diagnostic requirements. Complementary measurement methods based on coherent undulator radiation and dispersion scans are analyzed. A dedicated demodulation undulator is designed as a future platform for controlled studies, and measurements using the existing SXFEL seeding beamline are presented to show experimental signatures of energy modulation suppression.

The paper is organized as follows. Section~\ref{sec:2} outlines the principle of energy modulation and demodulation. Section~\ref{sec:3} presents numerical studies of the demodulation process. Section~\ref{sec:4} discusses diagnostics of residual energy modulation. Section~\ref{sec:5} describes the demodulation undulator design. Section~\ref{sec:6} presents preliminary experimental results at SXFEL. Finally, Section~\ref{sec:7} summarizes the main conclusions and perspectives.

\begin{figure}
\includegraphics[width=0.48\textwidth]{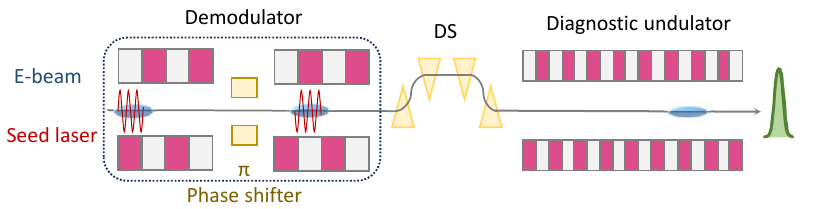}
\caption{Schematic layout of the energy modulation and demodulation in a seeded FEL. The demodulation undulator (demodulator) comprises two identical modulators separated by a tunable phase shifter. A diagnostic undulator is placed downstream of the dispersive section (DS) to amplify the coherent radiation emitted by the energy-modulated electron beam. }\label{fig:1}
\end{figure}

\section{\label{sec:2}Principle of energy modulation and demodulation}
Figure~\ref{fig:1} presents a schematic layout of an energy modulation–demodulation that closely resembles an externally seeded FEL, particularly the HGHG scheme \citep{Yu1991,Yu2002}. The system comprises two modulator undulators configured in tandem to form a demodulator, with a phase shifter symmetrically positioned between them. In a conventional HGHG setup, a UV seed laser imparts a sinusoidal energy modulation onto the electron beam as it passes through the modulator. This energy modulation is subsequently converted into a longitudinal density modulation (microbunching) via a downstream dispersive section.

In the present configuration, when the phase shifter is inactive (i.e., introduces no additional phase shift), the combined pair of undulators effectively functions as a single, extended modulator—thereby replicating the standard modulation stage of a seeded FEL. The dimensionless energy modulation amplitude $A$ and the output rms energy spread of the electron beam $\sigma_{\gamma}^{\prime}$ after the modulator can be expressed as:
\begin{equation}
A = \Delta \gamma/\sigma_{\gamma}
\end{equation}
\begin{equation}
\sigma_{\gamma}^{\prime}=\sqrt{\sigma_{\gamma}^{2}+\frac{\Delta \gamma^{2}}{2}}.
\end{equation}
\begin{figure}
\includegraphics[width=0.3\textwidth]{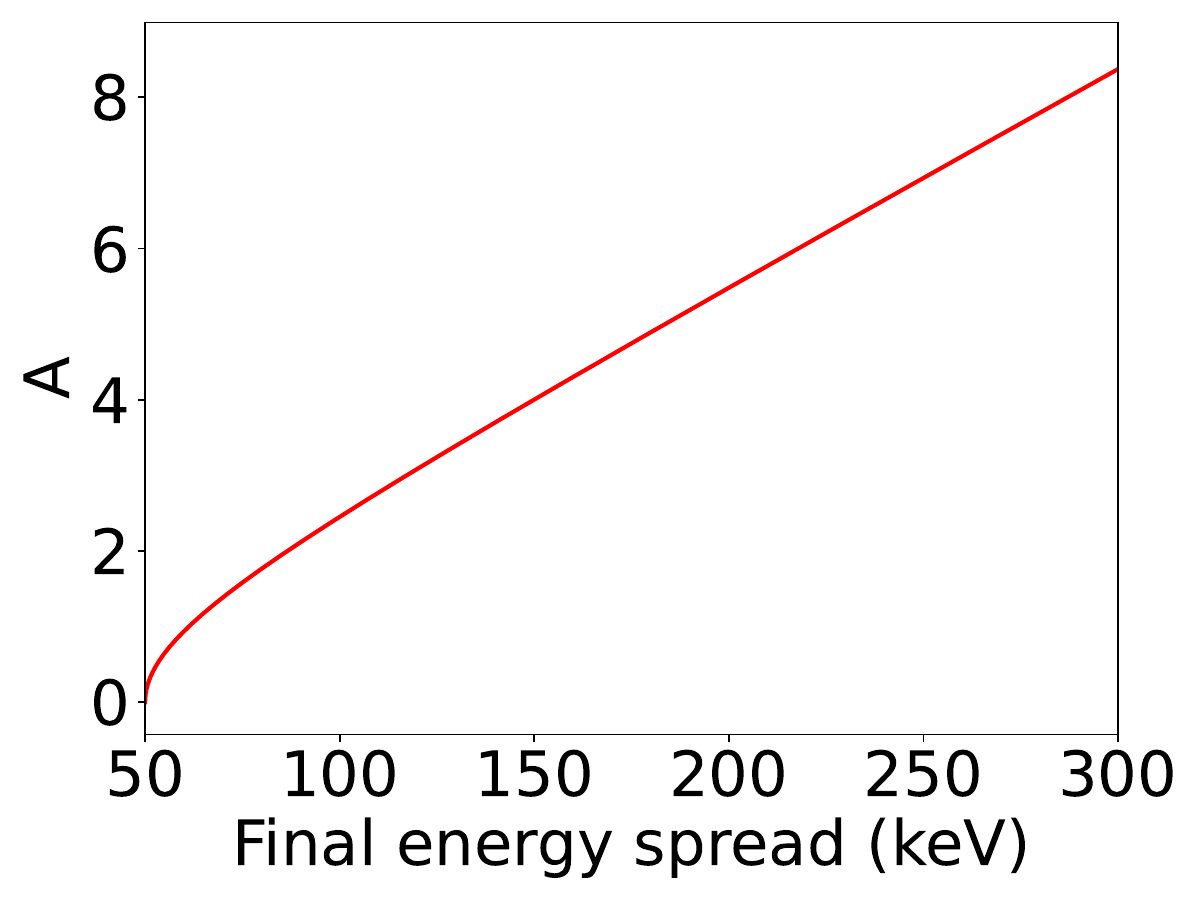}
\caption{Energy modulation amplitude versus rms energy spread relationship.}\label{fig:2}
\end{figure}

Thus, the final laser-induced energy modulation at the exit of the modulator is:
\begin{equation}
    A = \sqrt{ 2 \left( \left( \frac{\sigma_{\gamma}^{\prime}}{\sigma_{\gamma}} \right)^{2} - 1 \right) }.
    \label{eq:3}
\end{equation}
As shown in Fig.~\ref{fig:2}, the amplitude of the laser-induced energy modulation can be estimated from the final output rms energy spread. For example, an increase of 1 keV in the energy spread—from an initial value of 50 keV—corresponds to an energy modulation amplitude of 0.3.

Generally, assuming a Gaussian external seed laser interacts with the electron beam in the modulator undulator, we can obtain the amplitude of the energy modulation as (see, e.g., \cite{Huang2004})
\begin{equation}
  \Delta \gamma(r)=\sqrt{\frac{P}{P_{0}}} \frac{K[JJ]_{1} 
  L_{u}}{\gamma \sigma_{r}} \exp({-\frac{r^{2}}{4 \sigma_{r}^{2}}}),
    \label{eq:4}
\end{equation}
where $P$ is the peak power of the external seed laser, $P_0 = I_{A}mc^2/e \approx 8.7$ GW, $r$ is the radial position of the electron beam, and $\sigma_{r}$ is the rms laser spot size in the modulator undulator. The laser spot size $\sigma_{r}$ should be as comparable as possible to the electron beam size $\sigma_{x}$ to obtain the maximum energy modulation $\Delta \gamma(0)$.

Subsequently, a magnetic chicane is employed to achieve density modulation of the electron beam, which can be quantified by the bunching factor. The $n$-th harmonic bunching factor can be derived as:
\begin{equation}
    b_n=\left|J_n\left(nAB\right)\right|\exp{\left(-\frac{n^2B^2}{2}\right)}\label{eq:5}
\end{equation}
where $J_n$ is the $n$-th order Bessel function, $A = \Delta \gamma/\sigma_{\gamma}$, and $B = kR_{56}\sigma_{\gamma}/\gamma$ denote the dimensionless energy modulation amplitude and dispersion parameter, respectively. Here, $\gamma$ is the Lorentz factor, $\Delta \gamma$ represents the energy modulation amplitude induced by the seed laser, $\sigma_{\gamma}$ is the slice energy spread, $R_{56}$ is the dispersion strength of the chicane, and $k$ is the wavenumber of the seed laser. To achieve a sufficiently strong bunching factor at the $n$-th harmonic, it is generally desirable that $A$ exceed $n$. However, generating lasing at high harmonics presents a fundamental trade-off: while a large energy modulation (i.e., large $A$) is required to drive strong microbunching, excessive slice energy spread can suppress FEL gain. Specifically, when the relative energy spread $\sigma_{\gamma}/\gamma$ exceeds the FEL Pierce parameter, exponential amplification of the radiation is no longer sustained, leading to a significant reduction in output peak power \citep{Huang2007}.

When the diagnostic undulator is tuned to the $n$-th harmonic, the evolution of the CHG radiation is described by \citep{Yu2002}:
\begin{equation}
    P_{\mathrm{coh\ }}=\frac{Z_0\left(K\left[JJ\right]_1LIb_n\right)^2}{32\pi\sigma_x^2\gamma^2}\label{eq:6}
\end{equation}
where $Z_{0}$=~377$\Omega$ is the vacuum impedance, $K$ is the undulator parameter, $[JJ]_{1}$ is the planar undulator Bessel factor, $L$ is the radiator length, $b_{n}$ is the $n$-th bunching factor, $I$ is the peak current, and $\sigma_{x}$ refers the transverse beam size. CHG radiation is strongly coupled with the transverse beam size, the peak current, and the undulator length. While the bunching factor is normally associated with harmonic radiation performance in seeded FELs, it is used here mainly as a diagnostic observable. The intensity and/or spectral characteristics of the CHG radiation provide an indirect but sensitive means for measuring the slice energy spread and residual modulation amplitude \citep{Feng2013,Allaria2025}.

Furthermore, the electron beam first interacts with the seed laser in the initial modulator undulator, acquiring a sinusoidal energy modulation. After passing through the phase shifter, the seed laser introduces a $\pi$-phase delay, enabling reverse modulation of the electron beam and effectively canceling the initial energy modulation. We assume that laser-induced energy modulation in the demodulator involves two symmetrical processes and write the longitudinal phase space variables transformation for the passage through the demodulation undulators and phase shifter as follows:
\begin{equation}
p^{\prime} = p + A \sin \theta +A \sin (\theta+\phi), \quad \theta^{\prime} = \theta + p^{\prime}.
\end{equation}
Modulation to demodulation: $\phi = \pi + \Delta\phi$
\begin{equation}
p^{\prime}
%   &= p + 2A\cos\left(\theta + \frac{\Delta\phi}{2}\right)\sin\left(-\frac{\Delta\phi}{2}\right) \\
   = p - A\Delta\phi\cos\theta, \quad \Delta\phi\rightarrow 0,
\end{equation}
Then, the output energy modulation can be equivalent to:
\begin{equation}
A^{*} = A\Delta\phi\label{eq:09}
\end{equation}
Dual-modulation: $\phi = 2n\pi + \Delta\phi$
\begin{equation}
p^{\prime} = p + 2A\cos\left(\frac{\Delta\phi}{2}\right)\sin\theta,  \quad \Delta\phi\rightarrow 0
\end{equation}
Similarly, the energy modulation can be equivalent to:
\begin{equation}
A^{*} = 2A\cos\left(\frac{\Delta\phi}{2}\right).
\end{equation}

In the one-dimensional approximation, the phase shift $\Delta\phi$ arises from the total dispersion $R_{56}$ introduced by the demodulator, which can be expressed as $R_{56} = 2N\lambda_s$ \citep{mcneil2005,zhang2021}, where $N$ is the number of undulator periods and $\lambda_s$ is the resonant wavelength. For the electron beam, the phase variation induced by dispersion within the demodulator is given by:
\begin{equation}
\Delta\phi_i = k_s (\theta_i' - \theta_i) = \frac{2\pi}{\lambda_s} R_{56} \frac{\Delta\gamma_i}{\gamma_0},
\end{equation}
where $k_s = 2\pi/\lambda_s$ is the wavenumber, $\theta_i'$ and $\theta_i$ are the longitudinal positions of the $i$-th electron before and after the dispersive section, respectively, $\Delta\gamma_i$ is the energy modulation of the electron, and $\gamma_0$ is the average relativistic Lorentz factor of the beam.

The estimated average phase shift is thus:
\begin{equation}
\overline{\Delta\phi} = \frac{2\pi}{\lambda_s} \, R_{56} \, \frac{\overline{\Delta\gamma}}{\gamma_0}.\label{eq:13}
\end{equation}
This residual phase offset leads to a remaining energy modulation after the demodulation process. Notably, this phase shift is independent of the external seed laser wavelength. When the seed laser intensity is weak or the number of modulator periods is sufficiently small, the residual energy modulation following demodulation Eq.~\ref{eq:09} is likely to approach zero.

\begin{table}
	\caption{\label{tab:table1}Main simulated parameters of the SXFEL used for the numerical demodulation study.}
	\begin{ruledtabular}
    \begin{tabular}{lcc}
    Parameters				&Value	&Unit\\ \hline
    {\textbf{\textit{Electron beam}}}&		&\\
    Energy					&1.4		&GeV\\
    Slice energy spread 	&50     	&keV\\
    Normalized emittance  	&1	        &mm$\cdot$mrad\\
    Bunch charge			&500	    &pC\\
    Bunch length (FWHM)		&600    	&fs\\
    Peak current (Gaussian) &700	    &A\\
    {\textbf{\textit{Demodulator}}}&		&\\
    Period length                     &80		    &mm\\
    Period number                      &12$\times$2		\\
    {\textbf{\textit{Phase shifter}}}&		&\\
    Period length                     &80		    &cm\\
    Phase shift range                      &0-4$\pi$		&rad\\
    {\textbf{\textit{Seed laser}}}&		&\\
    Wavelength                     &266		    &nm\\
    Pulse duration (FWHM)                     &$~$0.1-1		&ps\\
    Peak power                             &$<$200       &MW\\
    Rayleigh length                              &$<$50       &m\\
    {\textbf{\textit{Diagnostic undulator}}}&		&\\
    Period length                    	&68     	&mm\\
    Length           	        &4	        &m\\
    {\textbf{\textit{Radiator undulator}}}&		&\\
    Period length                    	&30      	&mm\\
    Length          	        &3	        &m\\
    \end{tabular}
    \end{ruledtabular}
\end{table}

\section{\label{sec:3}Numerical studies of the demodulation process}
\begin{figure}
% \centering
\subfigure[\label{fig:3a}]{
\includegraphics[width=0.23\textwidth]{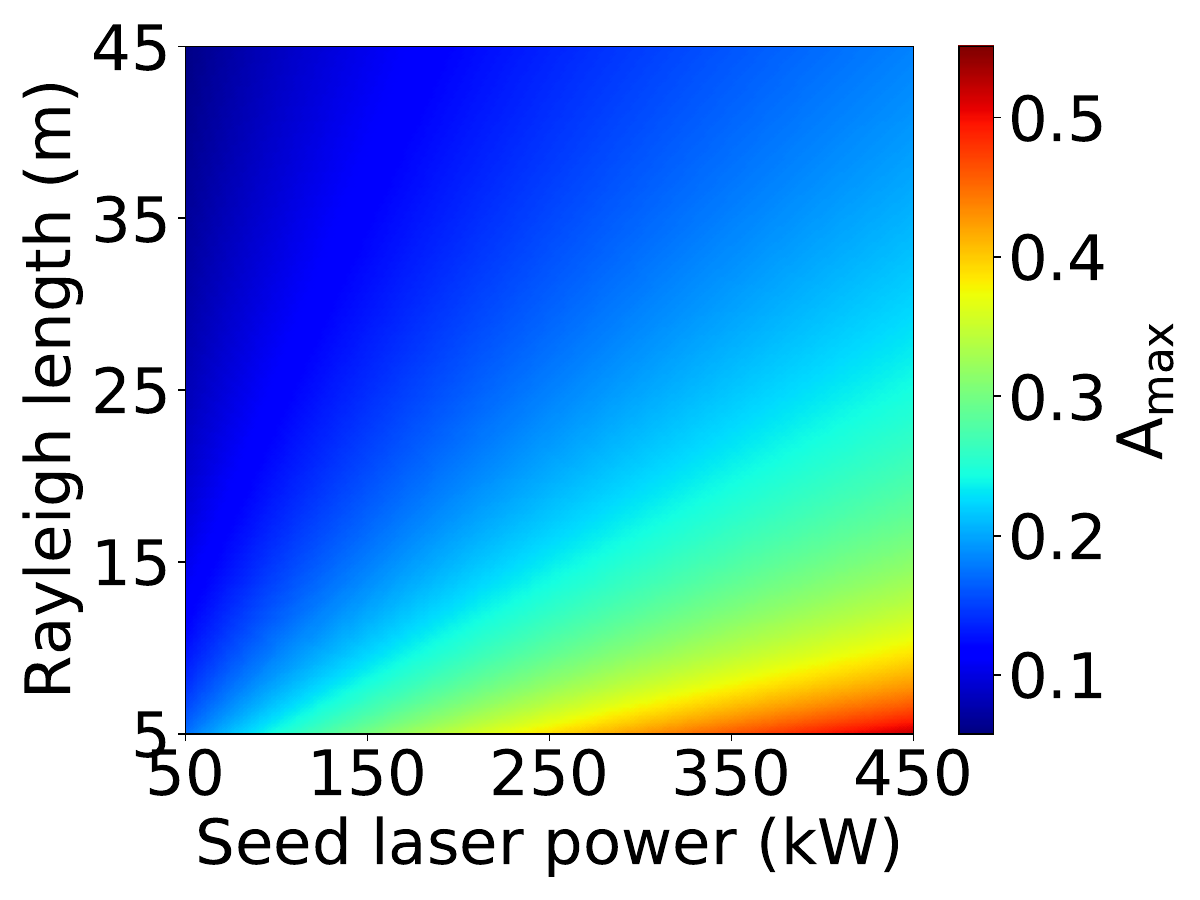}}
\subfigure[\label{fig:3b}]{
\includegraphics[width=0.23\textwidth]{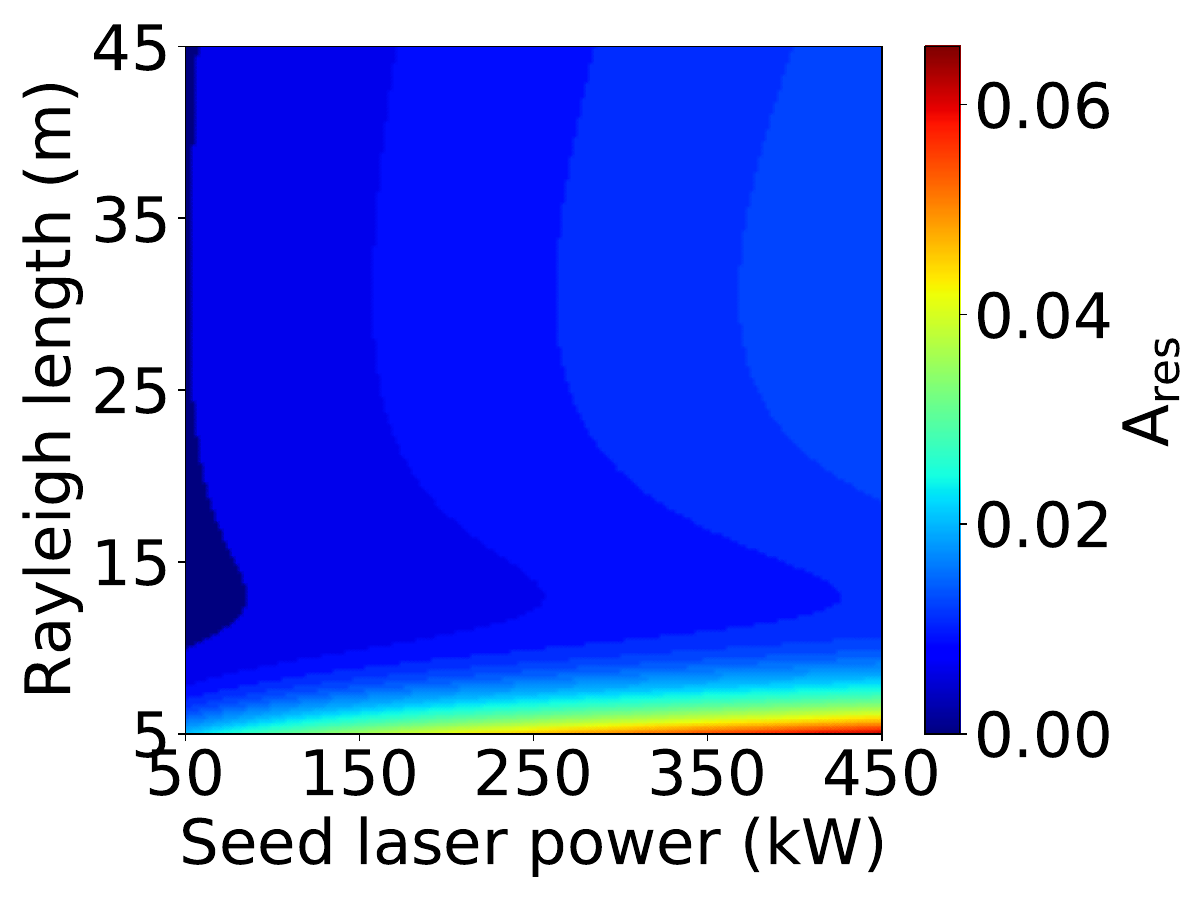}}
\subfigure[\label{fig:3c}]{
\includegraphics[width=0.23\textwidth]{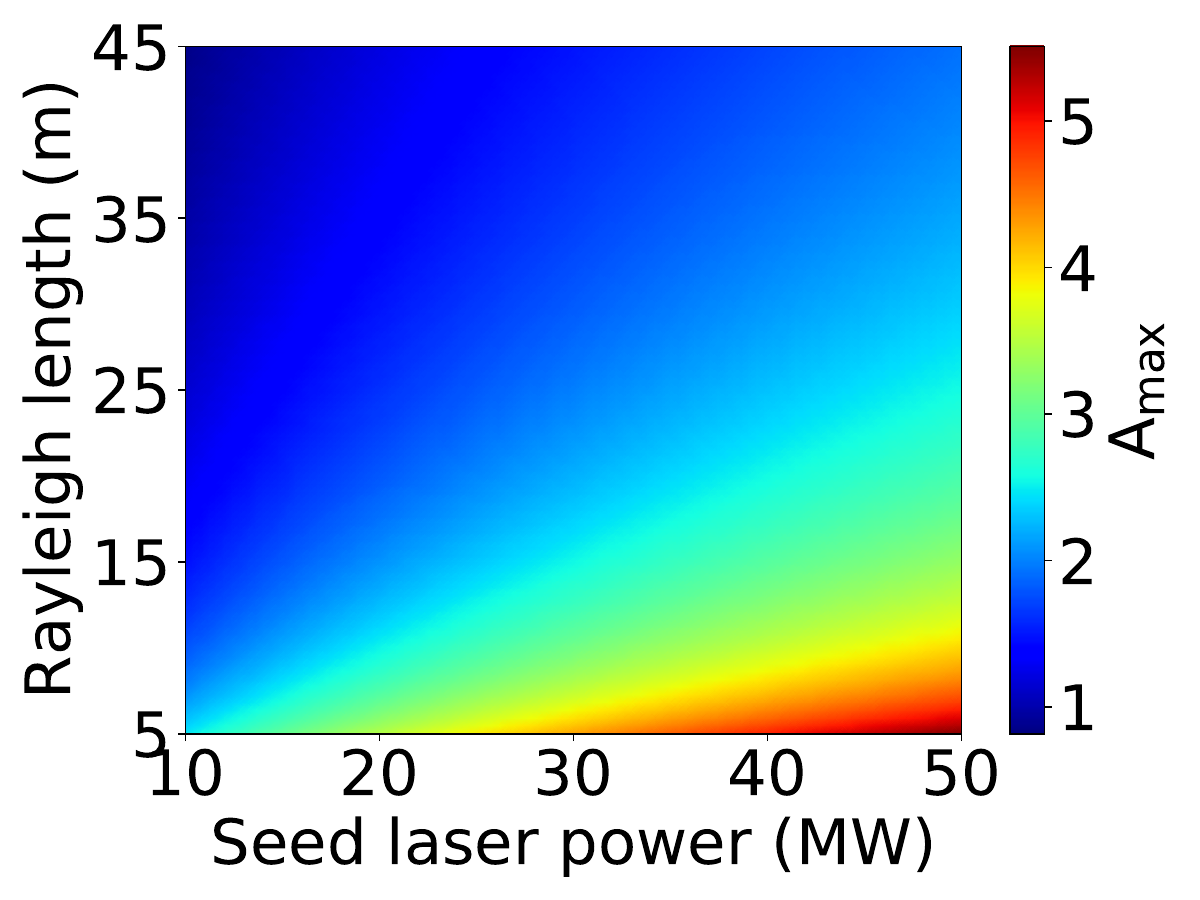}}
\subfigure[\label{fig:3d}]{
\includegraphics[width=0.23\textwidth]{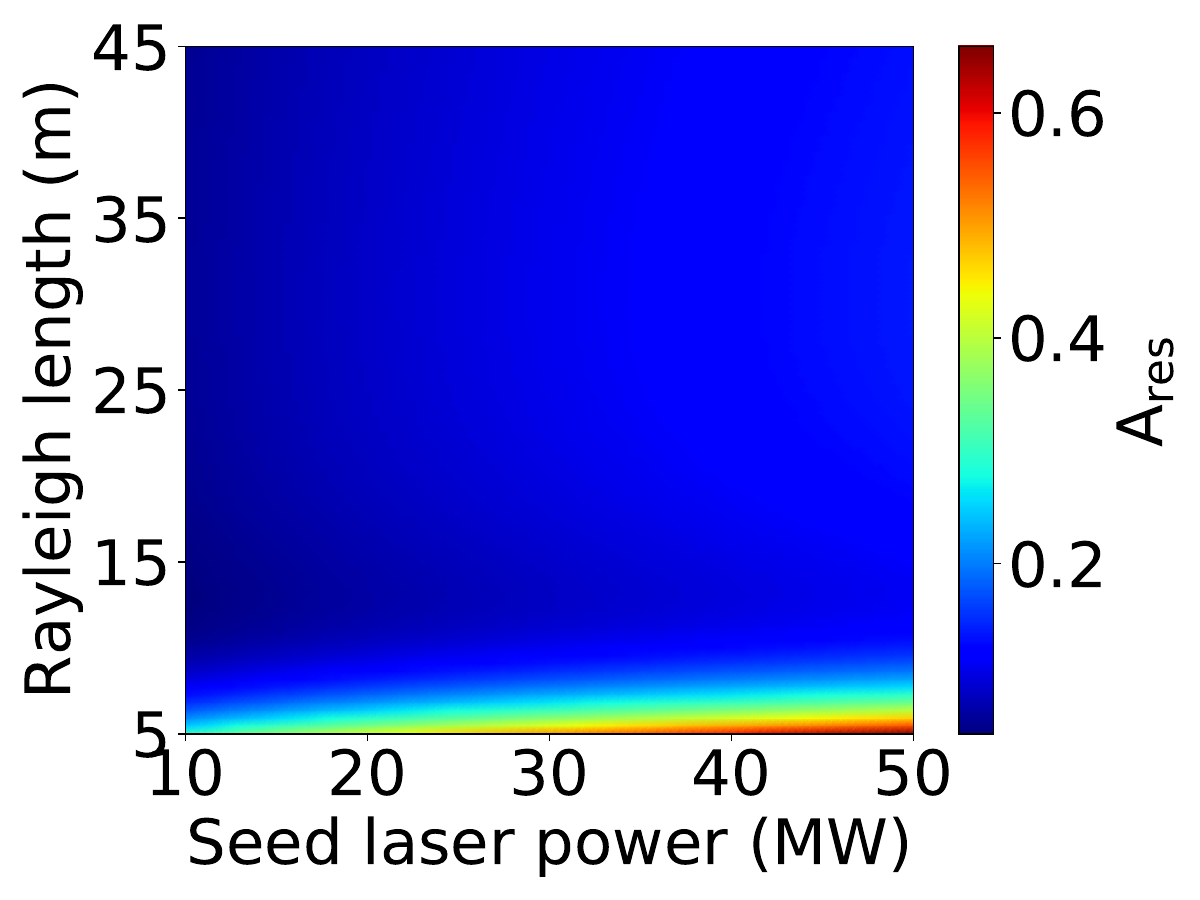}}
% \caption{Maximum energy modulation amplitude in panels (a) and (c), and residual energy modulation amplitude at the demodulator exit in panels (b) and (d), as functions of seed laser Rayleigh length and peak power in the demodulator, under conditions of weak and large initial energy modulation, respectively.}
\caption{Parameter scan of energy modulation and residual modulation in the demodulator. Panels (a, b) and (c, d) correspond to the weak- and large-modulation cases, respectively. Panels (a, c) show the maximum modulation amplitude $A_{\max}$, and panels (b, d) show the residual modulation amplitude $A_{\mathrm{res}}$ at the demodulator exit, as functions of seed-laser peak power and Rayleigh length.}
\label{fig:3}
\end{figure}

\begin{figure}
% \centering
\hspace*{-20pt}
\subfigure[\label{fig:4a}]{\includegraphics[width=0.23\textwidth]{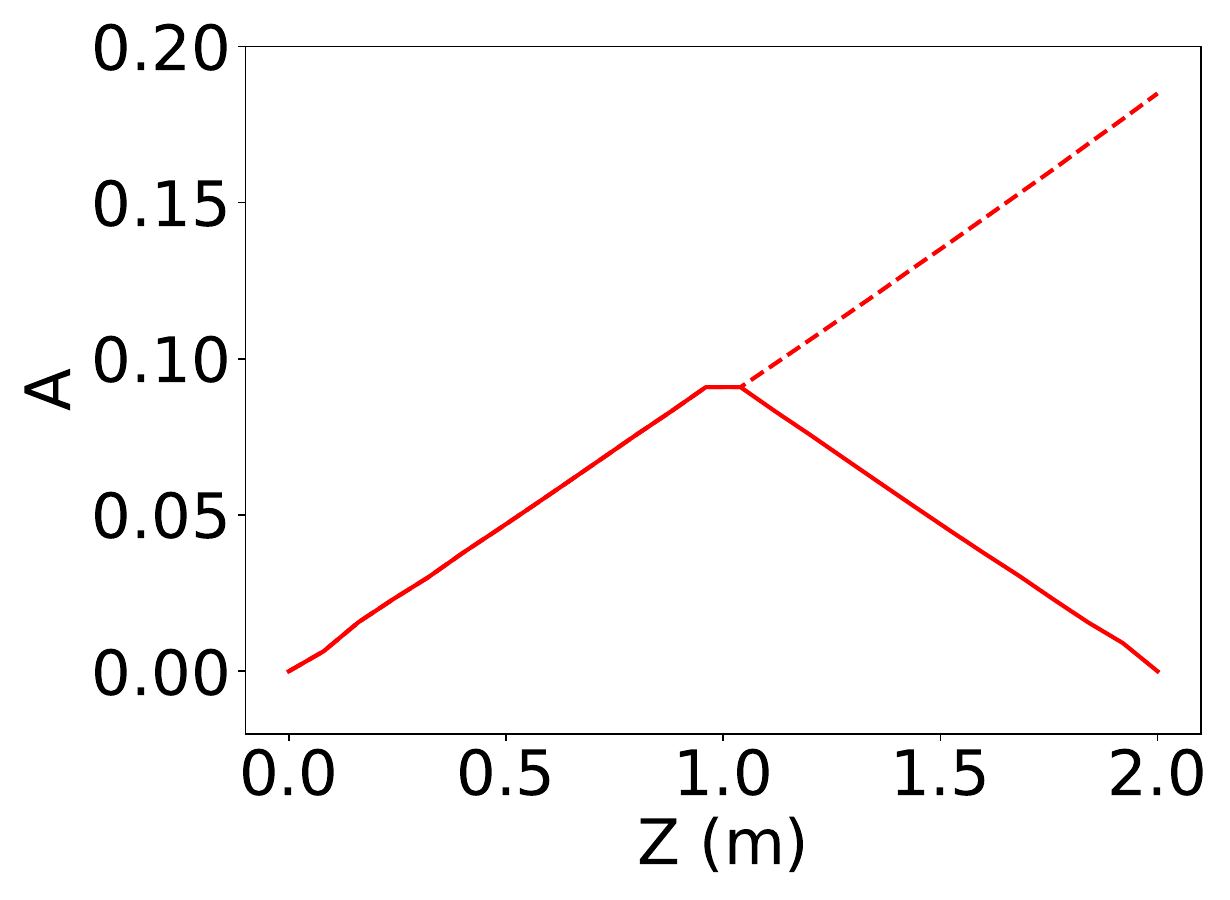}}
\subfigure[\label{fig:4b}]{\includegraphics[width=0.23\textwidth]{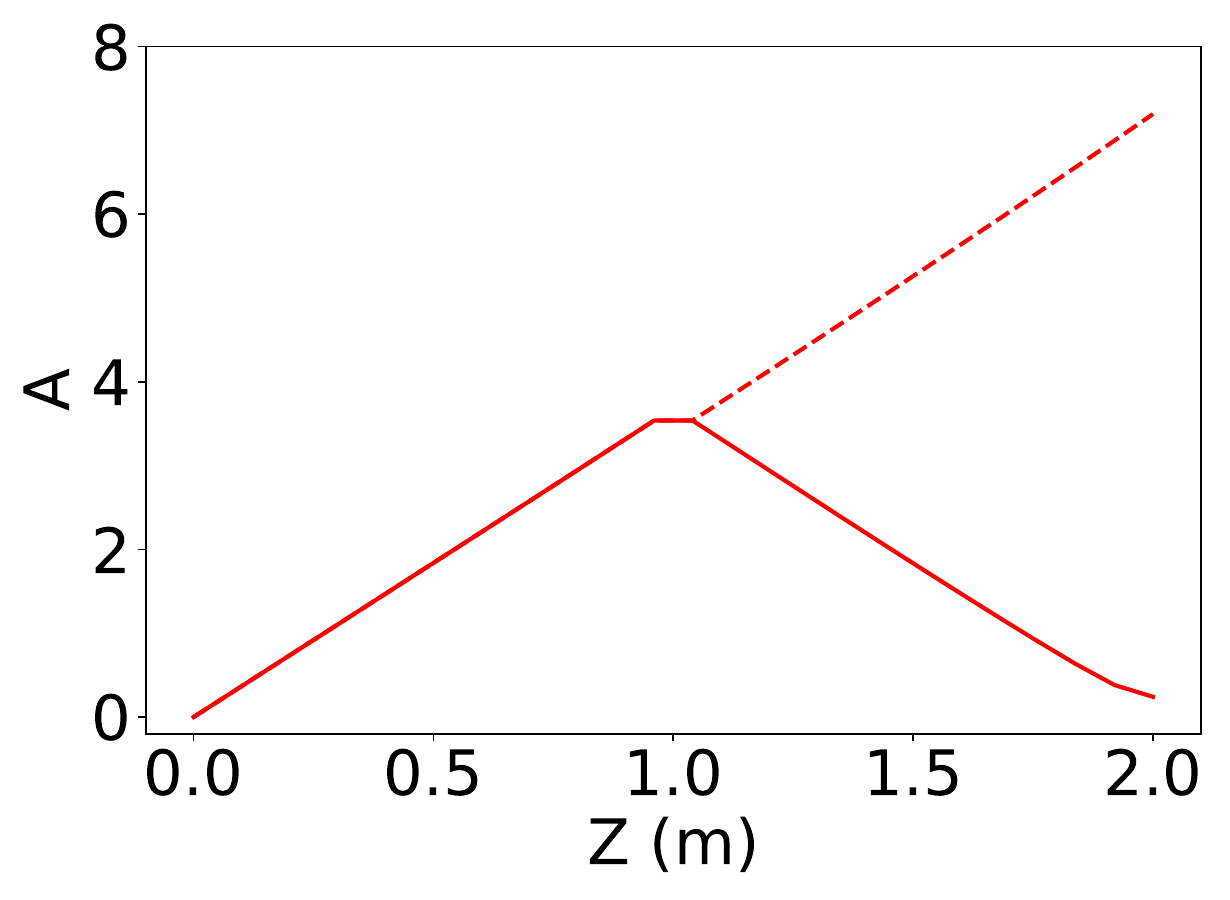}}
\hspace*{-20pt}
\subfigure[\label{fig:4c}]{ \includegraphics[width=0.23\textwidth]{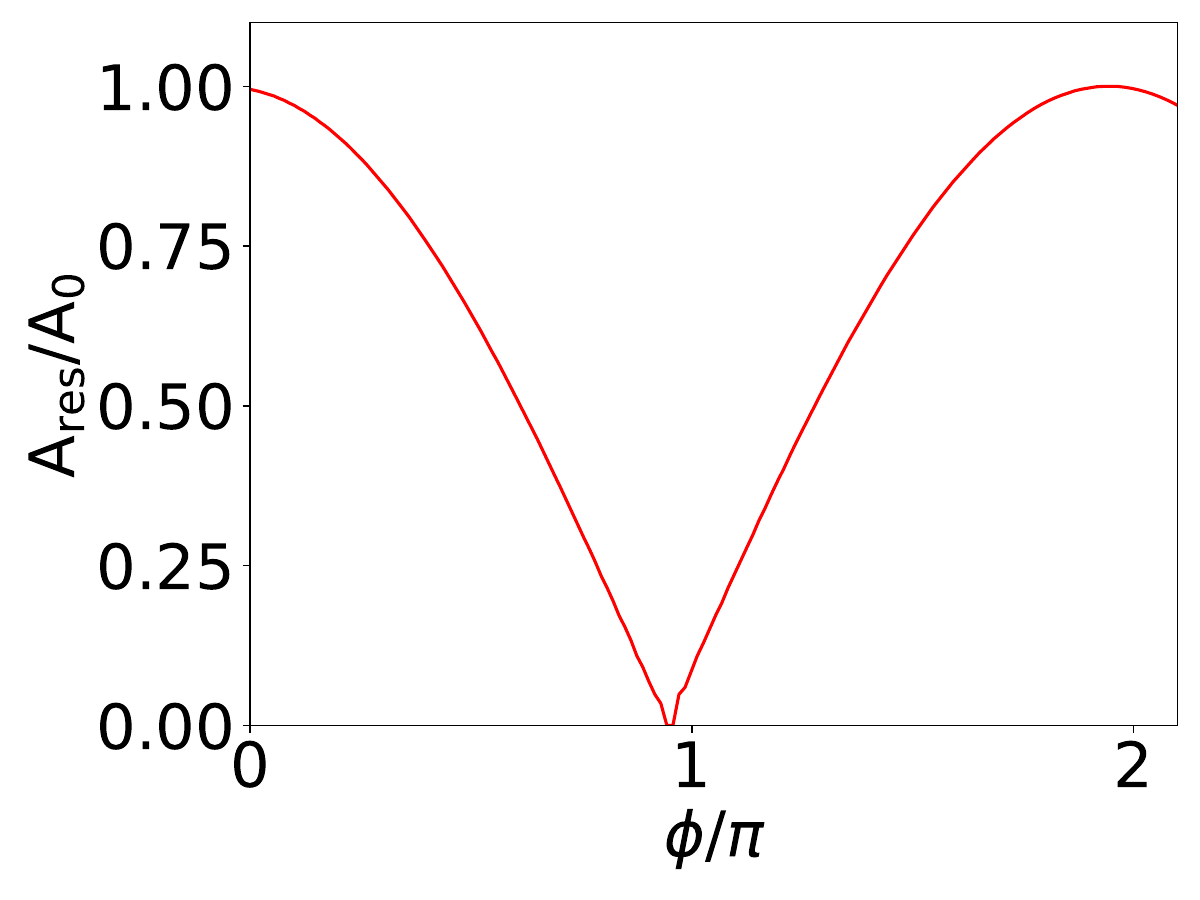}}
\subfigure[\label{fig:4d}]{ \includegraphics[width=0.23\textwidth]{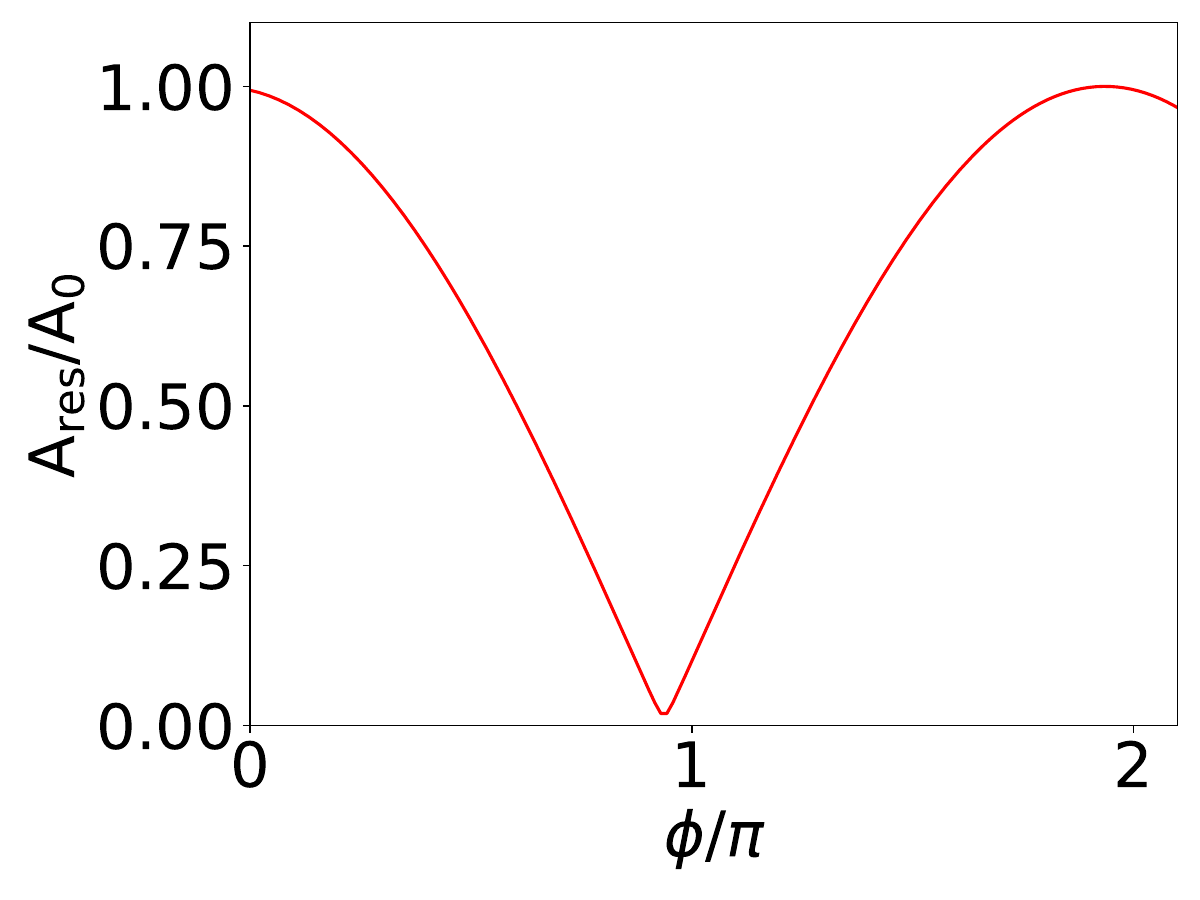}}
\caption{Evolution and phase dependence of the energy modulation in the demodulator.
Panels (a) and (b) show the evolution of the modulation amplitude along the demodulator for the weak- and strong-modulation cases, respectively. 
The dashed and solid curves correspond to the zero-phase and optimized phase-shift conditions.
Panels (c) and (d) show the suppression coefficient, $A_{\mathrm{res}}/A_0$, as a function of the phase shift for the weak- and large-modulation cases, respectively, where $A_0=A_{\mathrm{out}}(\phi=0)$ denotes the output modulation amplitude under the zero-phase condition.
A smaller value of $A_{\mathrm{res}}/A_0$ indicates stronger suppression of the laser-induced energy modulation.}
\label{fig:4}
\end{figure}

\begin{figure}
% \centering
\hspace*{-10pt}
{\includegraphics[width=0.48\textwidth]{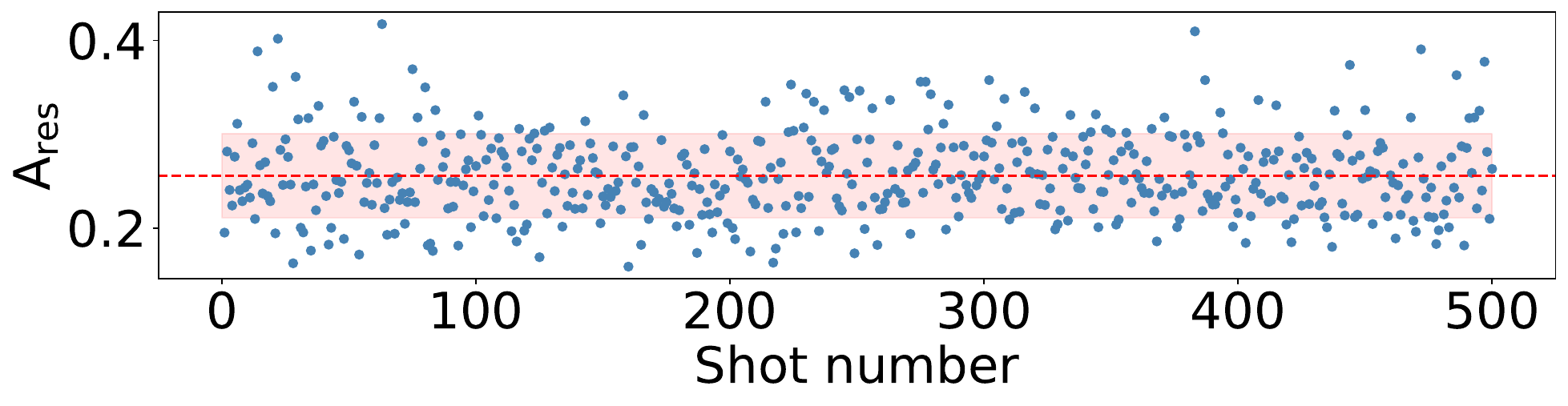}}
\caption{Shot-to-shot stability of the residual modulation amplitude $A_{\mathrm{res}}$ in the large-modulation case from 500 simulations including transverse beam-laser jitter. The dashed line and shaded region denote the mean value and rms fluctuation range, respectively.}
\label{fig:5}
\end{figure}

\begin{figure*}
\includegraphics[width=1\textwidth]{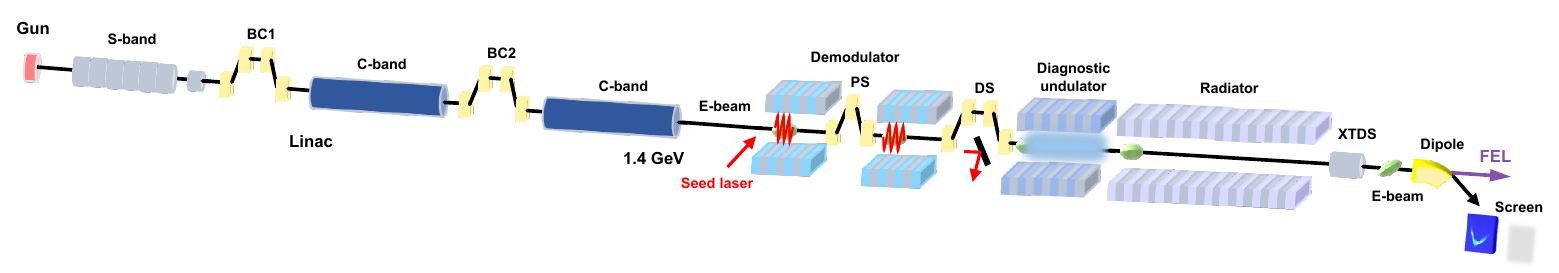}
\caption{Schematic illustration of the energy modulation and demodulation implemented at the SXFEL (not to scale).}\label{fig:6}
\end{figure*}

This section focuses on the numerical investigation of the electron-beam energy modulation--demodulation process. Within a simplified theoretical framework—excluding the radiation section and neglecting energy spread growth induced by radiation gain—we explore the fundamental limits of the demodulation process. As illustrated in Fig.~\ref{fig:1}, the basic layout consists of a demodulator composed of two identical modulator undulators, with a permanent-magnet phase shifter symmetrically positioned between them. To accurately examine the demodulation process, the lengths of the two modulators are restricted to minimize dispersion effects. A UV seed laser is employed for beam modulation, with parameters such as laser power and transverse beam size serving as key optimization variables. Besides, the SXFEL \citep{Liu2021} is uniquely capable of generating fully coherent radiation in the “water window” spectral range. The facility comprises a linac, two undulator lines, two beamlines, and six end-station experimental hutches. Based on these conditions, the main simulation parameters of the SXFEL employed as a representative example are summarized in Table~\ref{tab:table1}.

The three-dimensional simulations were performed with the GENESIS code \citep{REICHE1999} to investigate the steady-state interaction between relativistic electron beams and laser fields. The simulation employed two modulators, each with a period length of 80~mm and a limited number of periods ($N = 12$), driven by a seed laser with a wavelength of 266~nm. The position of the seed laser's waist is set at the geometric center of the demodulator to maintain symmetry in the energy modulation-demodulation process as much as possible. To quantify the modulation-suppression effect, we denote the residual modulation amplitude at the demodulator exit as $A_{\mathrm{res}}$. The zero-phase output modulation amplitude is defined as
$A_0 = A_{\mathrm{out}}(\phi=0)$. We further define the normalized residual modulation, or suppression coefficient, as
\[
\eta_s = \frac{A_{\mathrm{res}}}{A_0}.
\] As shown in Fig.~\ref{fig:3}, the maximum modulation amplitude $A_{\max}$ near the center of the demodulator and the residual modulation amplitude $A_{\mathrm{res}}$ at the exit were systematically studied by scanning the seed-laser peak power and Rayleigh length, under the optimal phase-shifter configuration. The modulation-suppression effect is quantified by the suppression coefficient $A_{\mathrm{res}}/A_0$. For the case with weak initial energy modulation, the optimal peak power of the seed laser is 50 kW, with a Rayleigh length of 20 m. In contrast, for strong initial energy modulation, the optimal peak power is 50 MW, with a Rayleigh length of 13 m.

Based on these conditions, the corresponding phase dependence of the suppression coefficient is shown in the lower panels of Fig.~\ref{fig:4}. This sharp minimum reflects the optimal demodulation condition, where the residual energy modulation is minimized. In the low-power seed case, the maximum energy modulation amplitude within the demodulator reaches 0.19, while at the exit, it is nearly zero, indicating successful demodulation. Although the demodulator introduces some longitudinal dispersion, the initial energy modulation is sufficiently weak that the electron beam's energy modulation profile remains close to an ideal sinusoidal distribution. In a more realistic scenario involving high-power seeding, the maximum energy modulation amplitude at the demodulator exit reaches 7.23, and the residual energy modulation at the exit is 0.14, corresponding to an energy spread of 50.3 keV, consistent with the value calculated from Eq.~(\ref{eq:13}), achieving approximately a 50-fold suppression in energy modulation amplitude. It is important to note that the stability of the seed laser parameters and the precision of the phase shifter scan significantly influence the demodulation performance. Moreover, measuring such small residual energy modulation amplitudes (below 0.3) is experimentally challenging. Therefore, we reduced the required scanning accuracy of the phase shifter. At a phase shift of $\phi = 0.96\pi$, the residual energy modulation amplitude is around 0.24.

To assess the sensitivity of the residual-modulation diagnostics to transverse laser-beam fluctuations, we performed 500-shot statistical simulations for the large-modulation case, as shown in Fig.~\ref{fig:5}. The electron beam has an rms transverse size of 100~$\mu$m, with 10\% rms size jitter; its injection position and angle exhibit rms jitters of 20~$\mu$m and 10~$\mu$rad, respectively. The seed laser has a FWHM spot size of 1.05~mm, a peak power of 50~MW, and rms fluctuations of 1\% in power and 5\% in spot size. Under these conditions, the residual modulation amplitude at the demodulator exit has an average value of $A_{\mathrm{res}}=0.26$ with an rms fluctuation of 17.5\%. The corresponding slice energy spread averages 50.8~keV, with an rms fluctuation of 0.6\%. These results indicate that the modulation-suppression effect remains observable at an experimentally accessible residual-modulation level under realistic transverse jitter.

The steady-state simulations in this section are intended to isolate the basic modulation-suppression mechanism and its diagnostic signatures. Additional longitudinal effects, such as relative timing jitter and seed-laser phase stability, can further affect the measurement of very weak residual modulation, but they do not change the underlying suppression mechanism. In practice, repeated phase scans, statistical averaging, improved laser-beam synchronization, and cross-checks using coherent undulator radiation and time-resolved diagnostics can improve the robustness of the measurement. Within the range of currently feasible experimental resolution, the following section describes the diagnostic procedures used to characterize the residual energy modulation and slice energy spread.

\section{\label{sec:4}Diagnostics of residual energy modulation}
The diagnostic study is based on the SXFEL seeding beamline infrastructure and modulator undulators, particularly those developed for the EEHG and cascaded EEHG-HGHG schemes \citep{Liu2021,Feng22,qi2025}. This configuration provides a realistic basis for evaluating residual energy modulation after the demodulation process, where the bunching factor is inferred from the coherent-radiation response of a short diagnostic undulator combined with dispersion scans rather than from FEL amplification at a target harmonic wavelength. A schematic of the complete system layout is presented in Fig.~\ref{fig:6}.

The diagnostic strategy is based on two complementary techniques, whose feasibility is validated through numerical simulations. Specifically, the coherent undulator radiation method \citep{Feng2013} is simulated using the \textsc{GENESIS} code, while the dispersion-scan method \citep{Prat2020a} is modeled with the \textsc{OCELOT} code \citep{Tomin2017}. Together, these simulations are employed to assess the resolution and robustness of the residual-modulation diagnostics. This section focuses on the high-power seed laser case, where a phase shift of $\phi = 0.96\pi$ yields a residual energy modulation amplitude of approximately 0.24—within the range of current experimental detectability. These techniques are strategically deployed in an alternating configuration based on specific measurement requirements. This dual-diagnostic approach enables comprehensive cross-validation of experimental results, yielding both indirect assessment of modulation fidelity during the FEL amplification process and direct time-resolved characterization of the electron beam's final energy distribution. Notably, this similar methodology parallels the diagnostic framework implemented at FERMI \citep{Allaria2025}, demonstrating the universality of this approach for seeded FEL characterization.

\subsection{\label{sec:4a}Determination of initial slice energy spread and modulation amplitude}
Accurate measurement of the energy modulation amplitude requires precise characterization of the electron beam’s initial slice energy spread. Based on Eqs.~(\ref{eq:5} and \ref{eq:6}), the modulation amplitude $A$ is inferred indirectly from the measured energy modulation $\Delta\gamma$ and the dispersion strength $R_{56}$, via their established relationship. This section focuses on the high-power seed laser case, where a phase shift of $\phi = 0.96\pi$ yields a residual energy modulation amplitude of approximately 0.24—within the range of current experimental detectability.

\begin{figure}
% \centering
\hspace*{-20pt}
\subfigure[\label{fig:7a}]{
\includegraphics[width=0.235\textwidth]{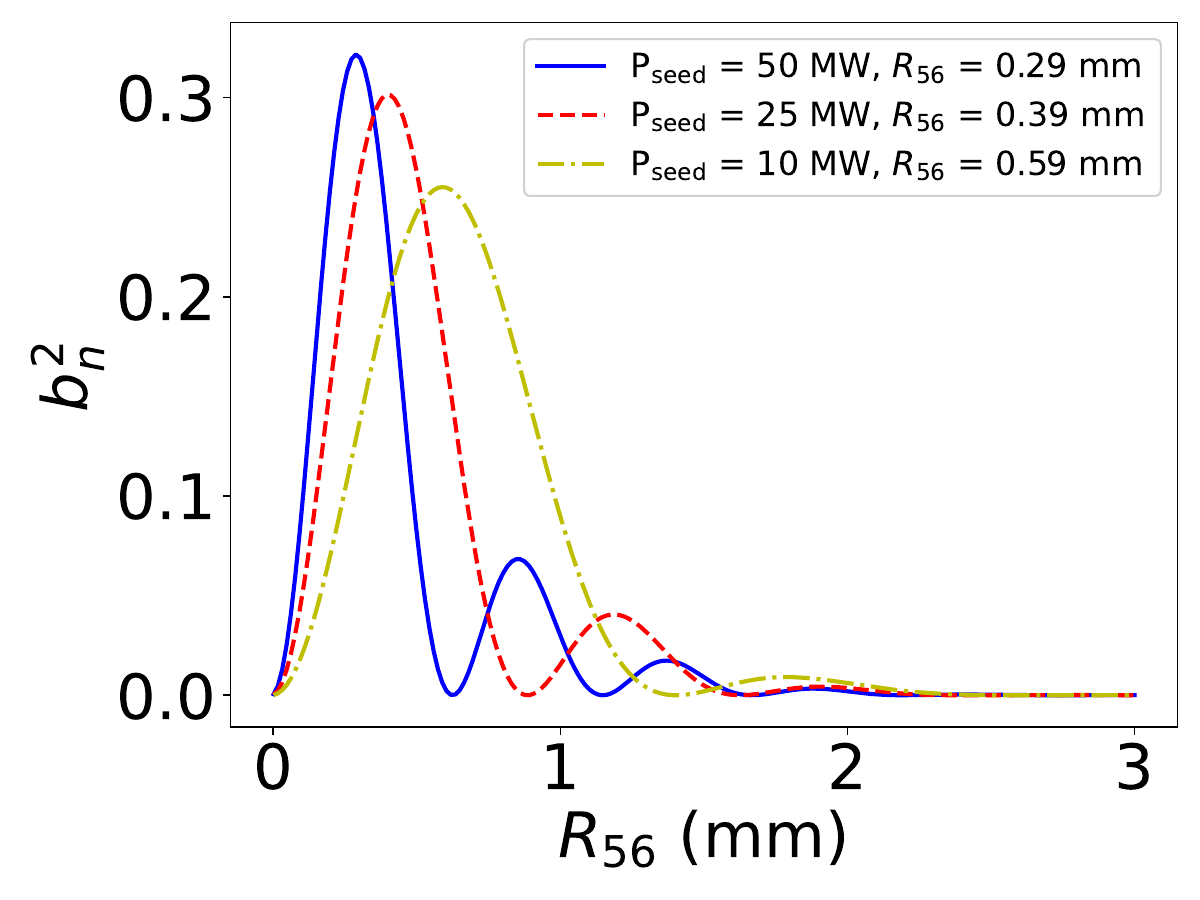}}
\subfigure[\label{fig:7b}]{
\includegraphics[width=0.235\textwidth]{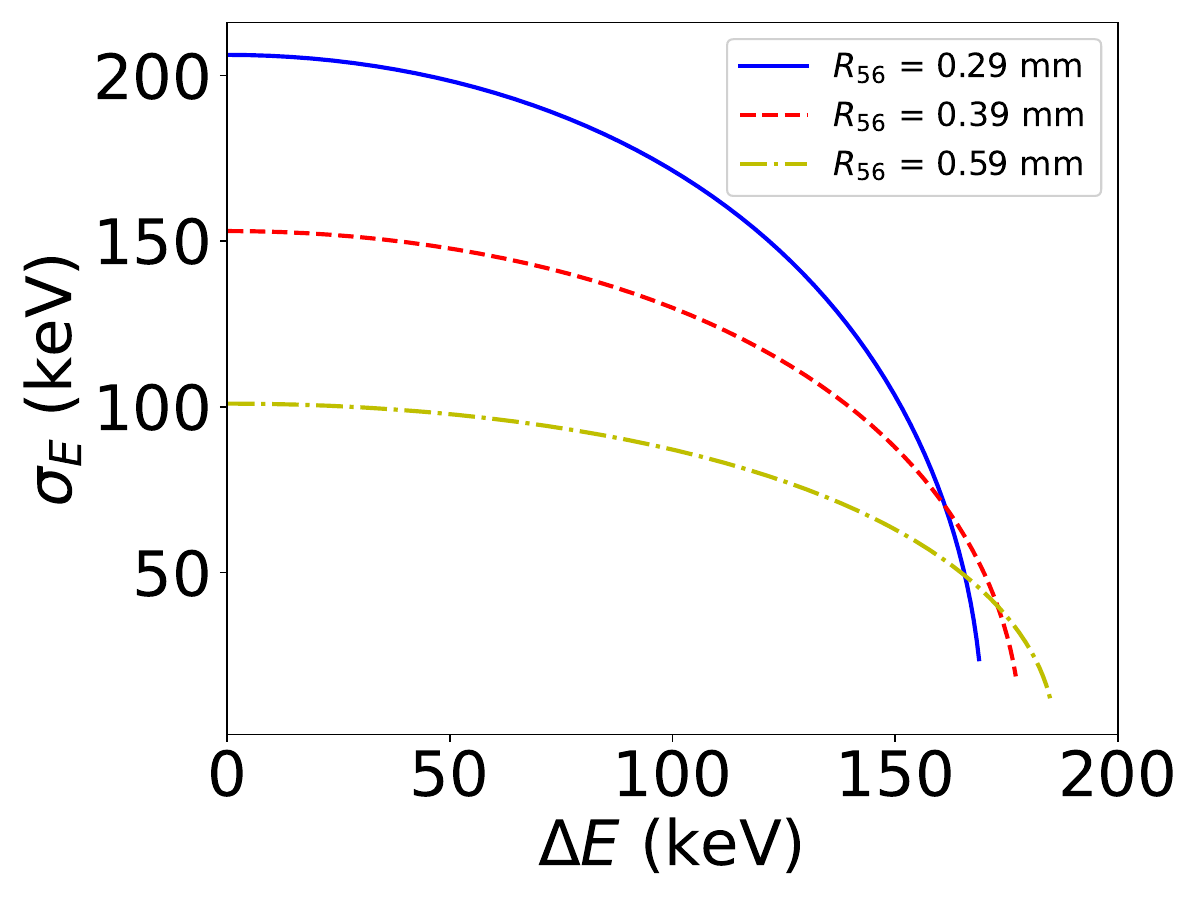}}
\hspace*{-20pt}
\subfigure[\label{fig:7c}]{
\includegraphics[width=0.235\textwidth]{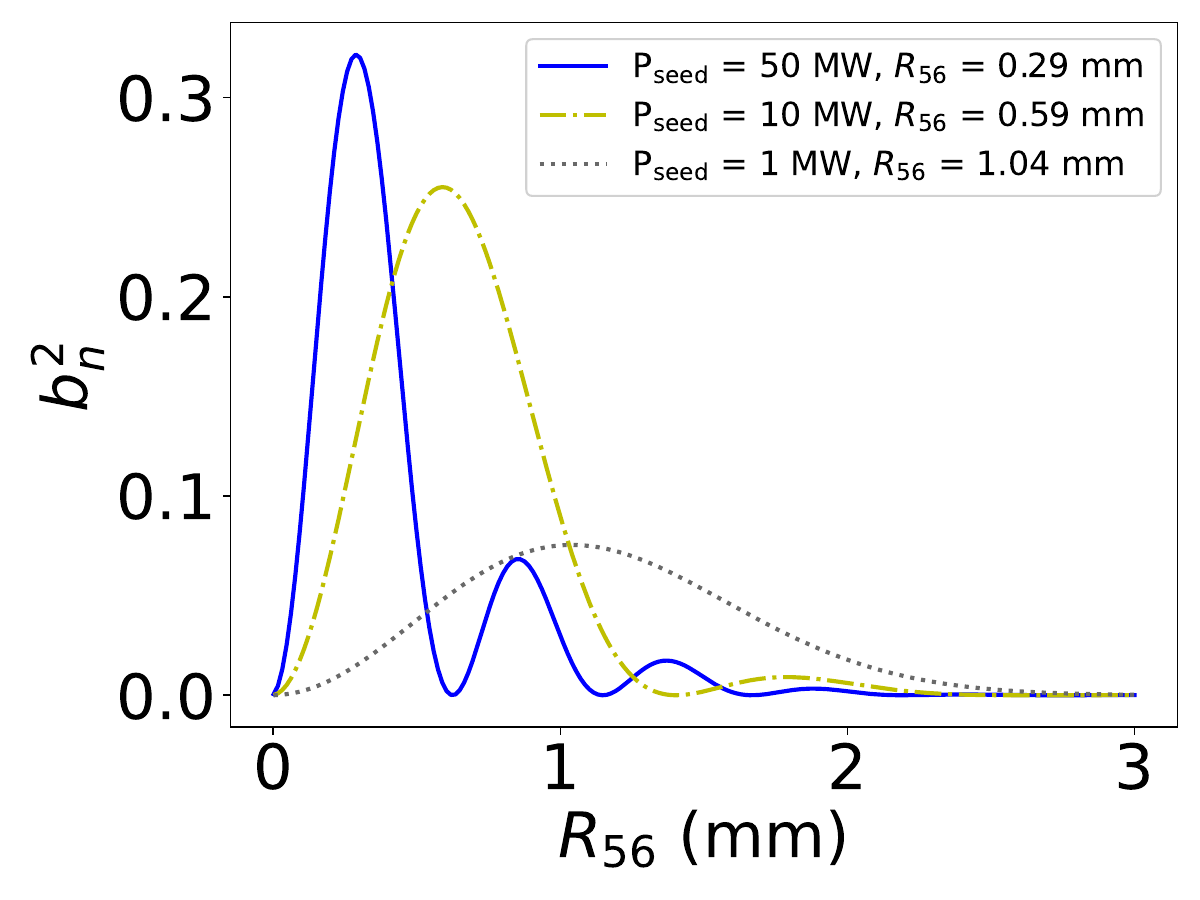}}
\subfigure[\label{fig:7d}]{
\includegraphics[width=0.234\textwidth]{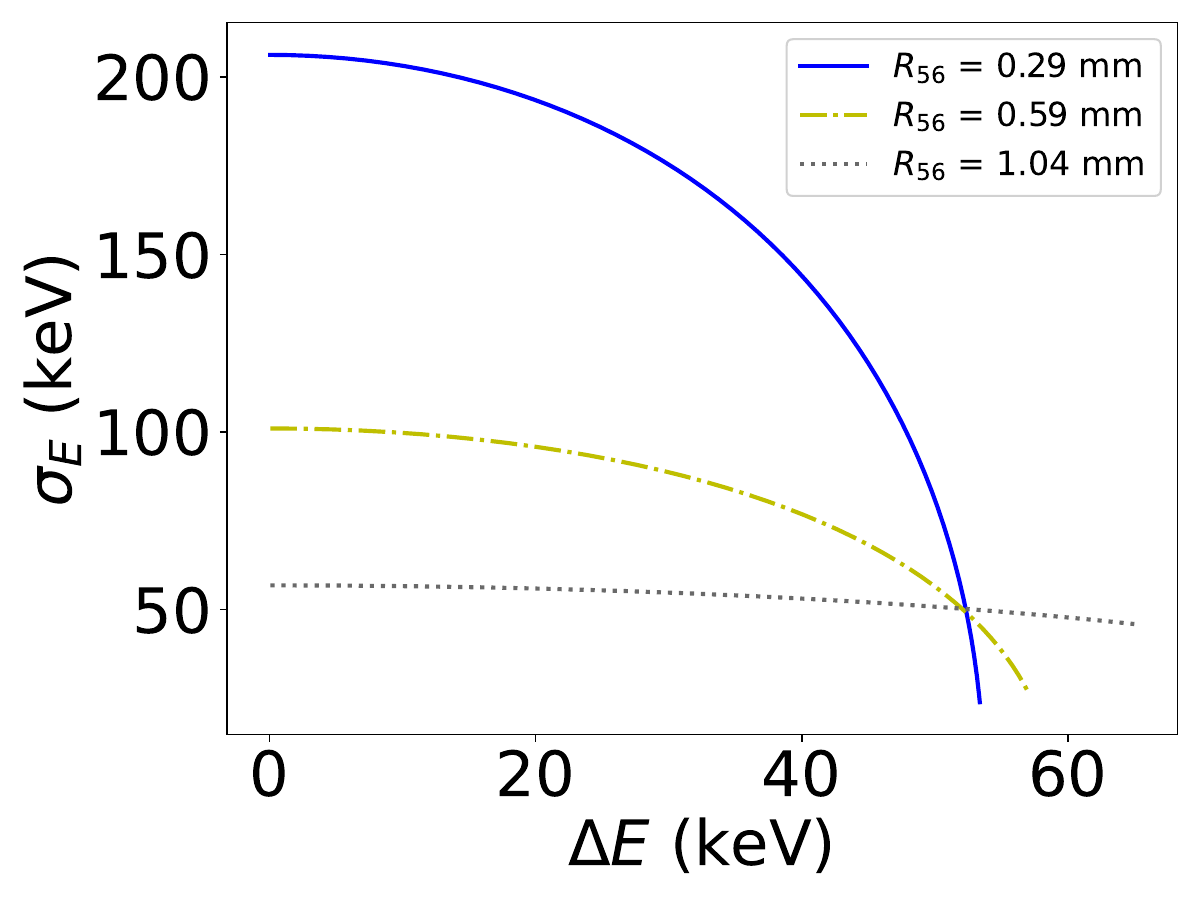}}
\caption{Simulated bunching factor squared as a function of dispersion strength under different seed laser power (a) and (c), and calculation results of slice energy spread and energy modulation amplitude (b) and (d).}
\label{fig:7}
\end{figure}

The procedure involves varying the seed laser intensity, scanning the dispersion strength $R_{56}$, and recording the corresponding intensity changes of the coherent undulator radiation at the target wavelength. The relationship between the optimal $R_{56}$ value and the energy modulation amplitude is then used to determine the energy spread and modulation characteristics \citep{Feng2013}. 

We utilize a diagnostic undulator with a period length of 68 mm and a total length of 4 m to characterize the electron beam's slice energy spread and energy modulation. The phase shifter is first tuned to maximize the fundamental coherent radiation intensity at the undulator exit; this condition corresponds to peak energy modulation and defines the zero-phase reference ($\phi = 0$). With the phase shifter fixed at this optimum, the dispersion strength $R_{56}$ is scanned to record the variation of the coherent radiation signal intensity. By performing measurements with multiple seed laser power settings that differ significantly, and ensuring that at least one case corresponds to a relatively low-power seed laser inducing an energy modulation amplitude of approximately one, a higher accuracy in the determination of the energy spread can be achieved. From the resulting curves, both the initial slice energy spread and the maximum energy modulation amplitude at zero phase are extracted.

As shown in Figs.~\ref{fig:7a} and ~\ref{fig:7c}, four seed laser powers are employed: 50~MW, 25~MW, 10~MW, and 1~MW, which are grouped into two sets for analysis. In Figs.~\ref{fig:7b} and ~\ref{fig:7d}, each set yields three intersection points among the $R_{56}$-scan curves at the known optimal dispersion strength, corresponding to three independent estimates of the slice energy spread. For the first group, the inferred values are 70.3~keV, 49.1~keV, and 39.9~keV, yielding an average of 53.0~keV—deviating by 3~keV from the true value of 50 keV. In contrast, the second group gives 50.2~keV, 50.1~keV, and 49.1~keV, with an average of 49.8~keV, deviating by only 0.2~keV from the true slice energy spread. The corresponding estimate of the maximum energy modulation amplitude induced by the seed laser is 7.40, in excellent agreement with the true value of 7.23.
\begin{figure}
% \centering
\hspace*{-20pt}
\subfigure[\label{fig:8a}]{
\includegraphics[width=0.235\textwidth]{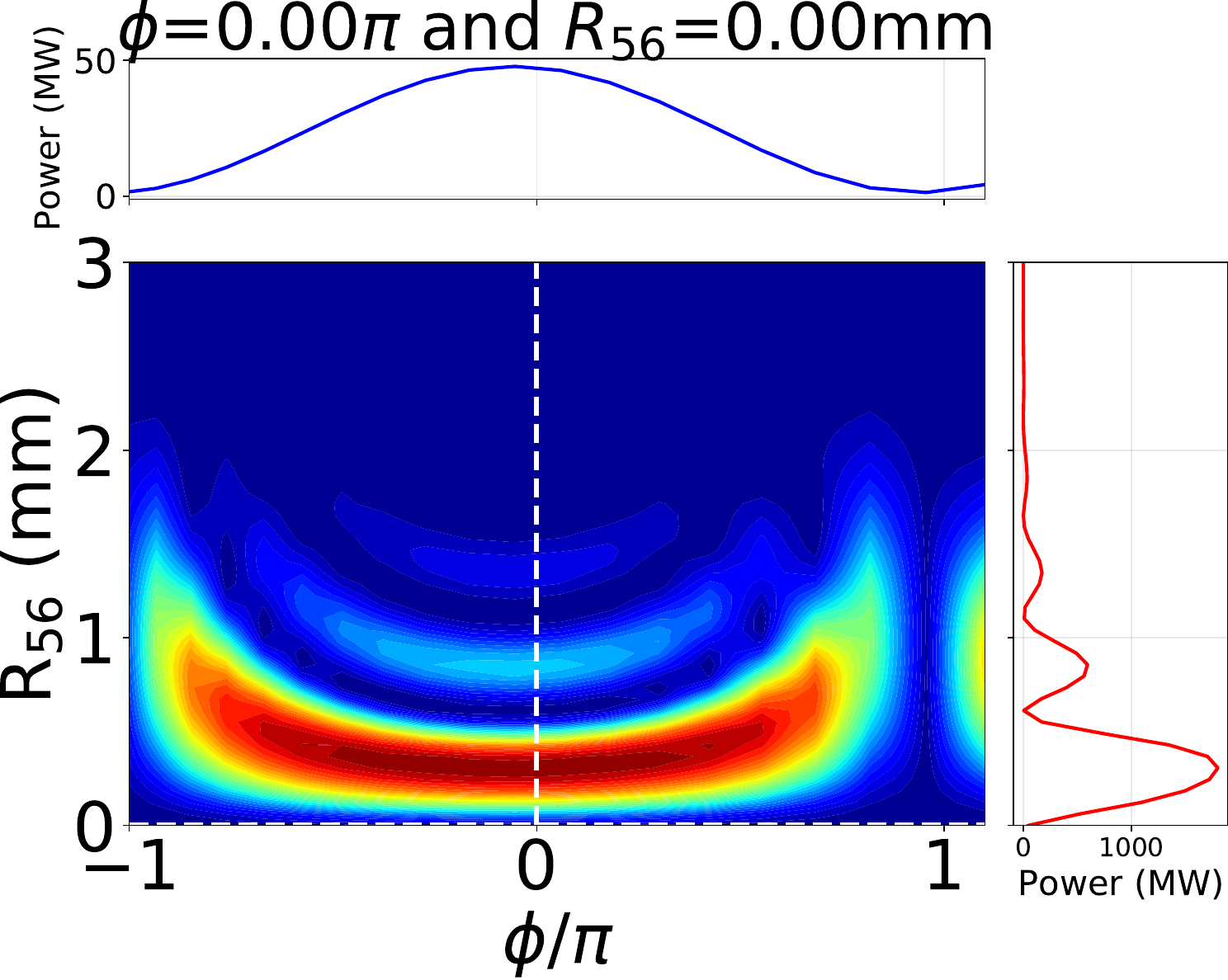}}
\subfigure[\label{fig:8b}]{
\includegraphics[width=0.235\textwidth]{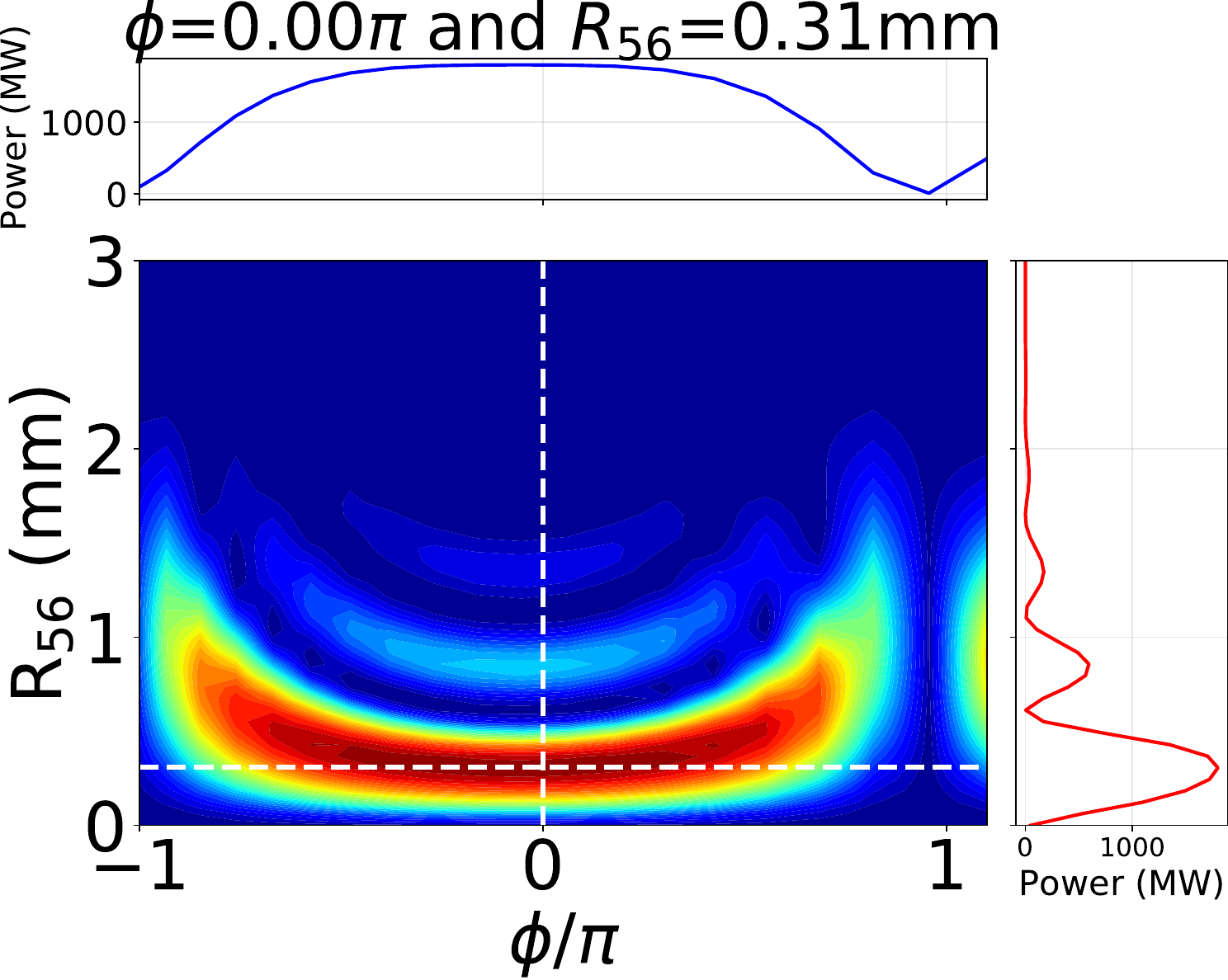}}
\hspace*{-20pt}
\subfigure[\label{fig:8c}]{
\includegraphics[width=0.235\textwidth]{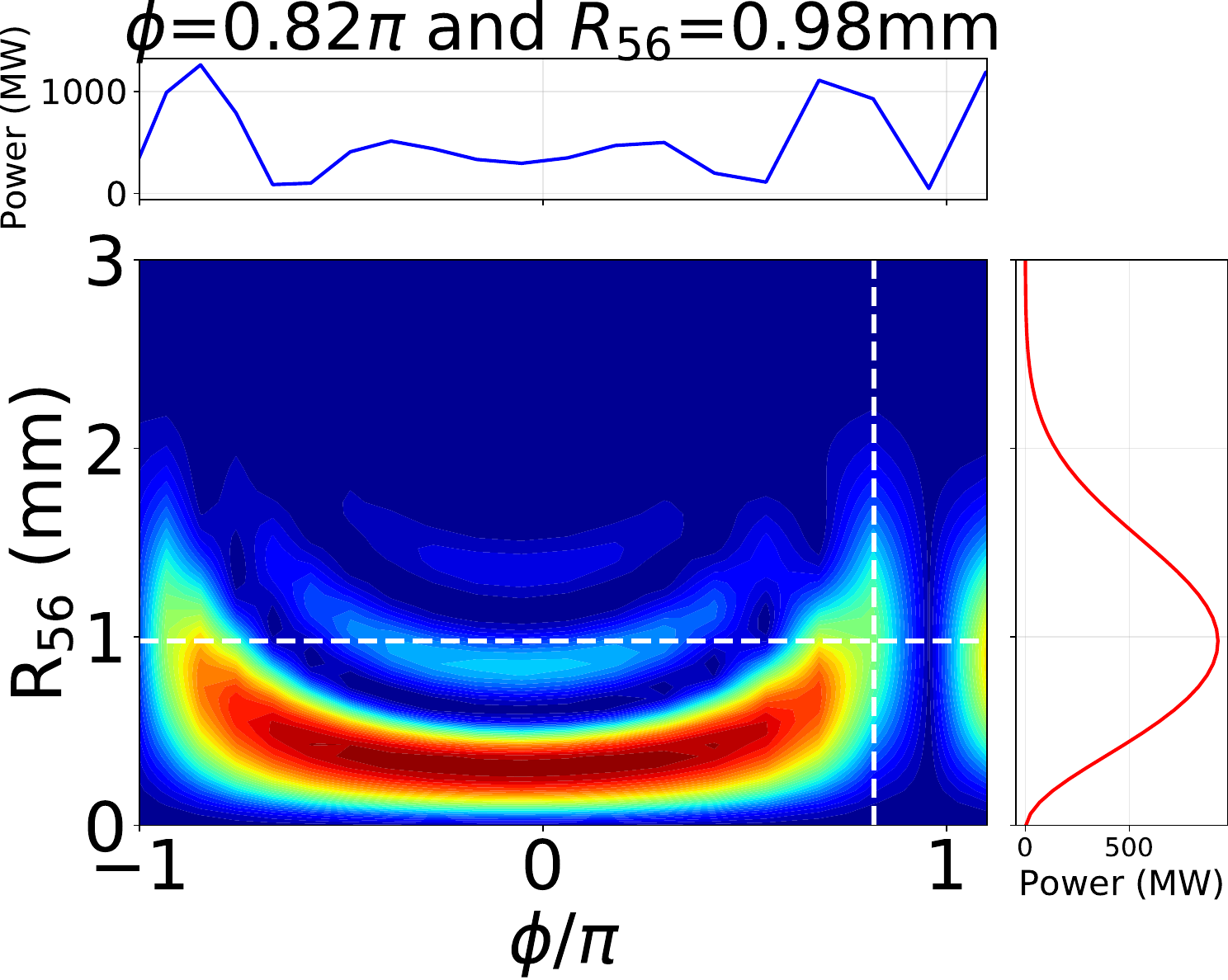}}
\subfigure[\label{fig:8d}]{
\includegraphics[width=0.235\textwidth]{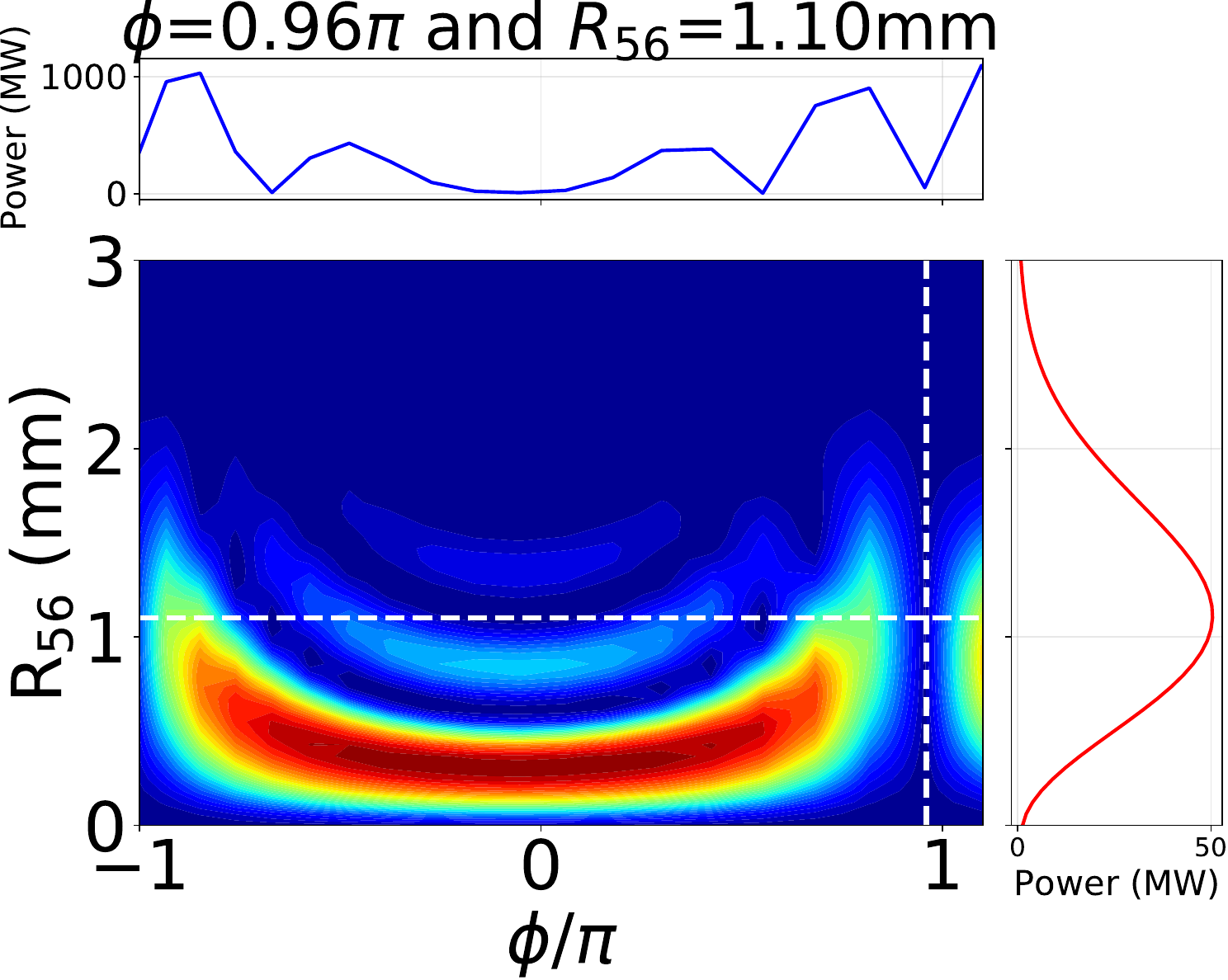}}
\caption{Coherent radiation intensity downstream of the diagnostic undulator as a function of the phase-shift $\phi$ and the $R_{56}$ value of the dispersive section, used to diagnose weak residual energy modulation. The white dashed lines represent the projected profiles for various phase-shift $\phi$ and $R_{56}$ values.}
\label{fig:8}
\end{figure}

\subsection{\label{sec:4b}Signatures of the demodulation process}
Variations in the energy modulation can be reflected by changes in the optimal dispersive strength $R_{56}$ and can therefore be used for indirect measurements. 
In addition, the evolution of the coherent radiation intensity measured downstream of the diagnostic undulator provides a fast and qualitative description of the coherent energy modulation--demodulation process.

At the initial working point ($\phi = 0$, $R_{56} = 0$), the electron beam is already modulated by a 50~MW seed laser, corresponding to a true energy modulation amplitude of 7.23. As shown in Fig.~\ref{fig:8a}, without introducing additional dispersion, the modulated electron beam directly enters the diagnostic undulator, where the coherent radiation signal is amplified by the intrinsic $R_{56}$ of the undulator itself. When the phase shifter is scanned over a $4\pi$ range, the radiation intensity exhibits a pronounced sinusoidal-like dependence on the phase shift $\phi$, which directly indicates the occurrence of the energy modulation--demodulation process. As $R_{56}$ is varied, the coherent radiation intensity exhibits pronounced sharp spikes near odd multiples of $\pi$, as shown in Fig.~\ref{fig:8b}. The inverted spike observed in the phase-shift--$R_{56}$ scan corresponds to a sharp local minimum in the coherent radiation intensity downstream of the diagnostic undulator. Figure~\ref{fig:8} shows that this feature arises from the strong sensitivity of the projected observable to the longitudinal phase-space evolution near the optimal demodulation condition. At this point, the residual energy modulation is minimized, leading to nearly complete destructive interference in the coherent radiation process. 

Notably, during the phase-shift $\phi$ scan, a well-defined minimum persists, whose location directly identifies the optimal demodulation phase. As a result, even a small deviation in either the phase shift or the $R_{56}$ value produces a rapid increase in the radiation intensity, giving rise to a cusp-like minimum in the projected intensity distribution. Such inverted spikes therefore provide a sensitive signature of the optimal demodulation condition and enable precise diagnostics of weak residual energy modulation.

\subsection{\label{sec:4c}Determination of residual energy modulation}
\begin{figure}
\centering
% \hspace*{-20pt}
\subfigure[\label{fig:9a}]
{\includegraphics[width=0.35\textwidth]{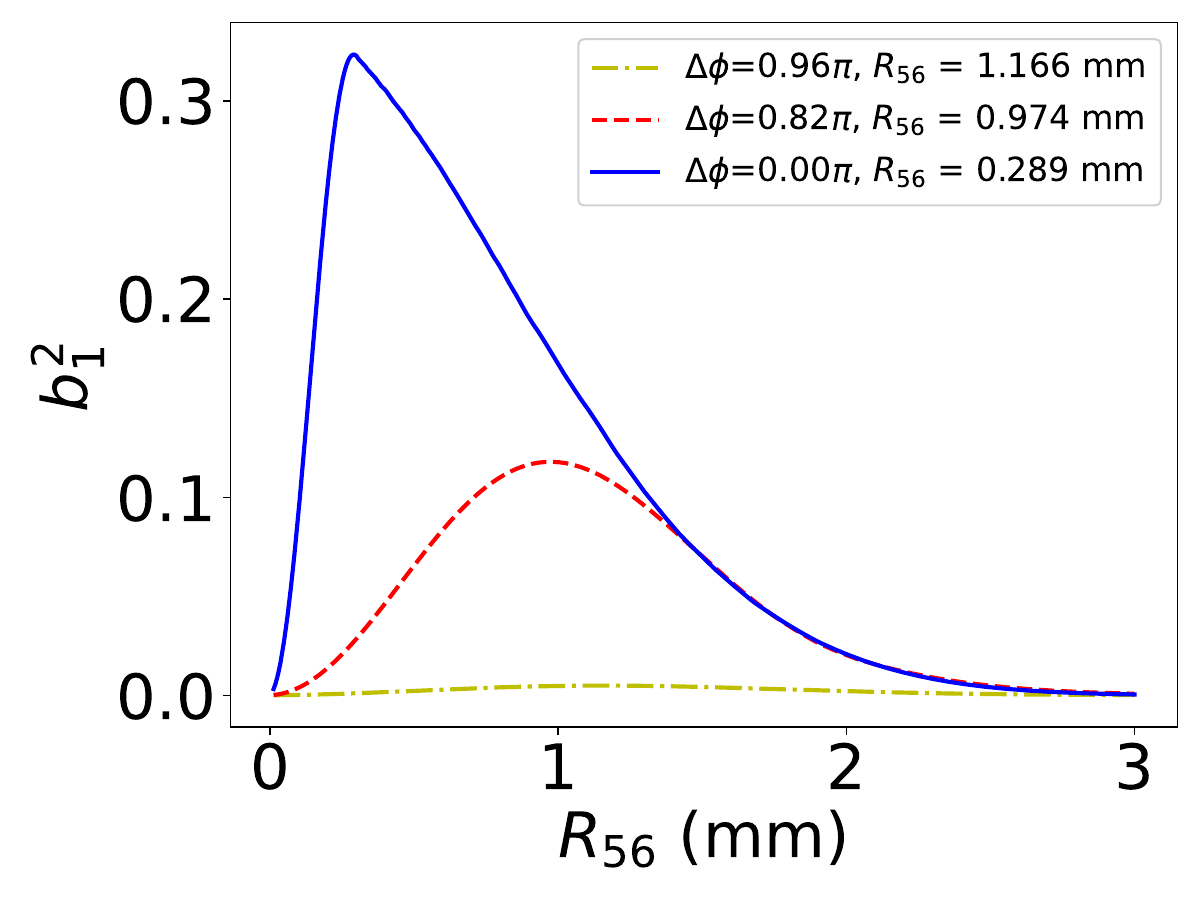}}
\subfigure[\label{fig:9b}]
{\includegraphics[width=0.35\textwidth]{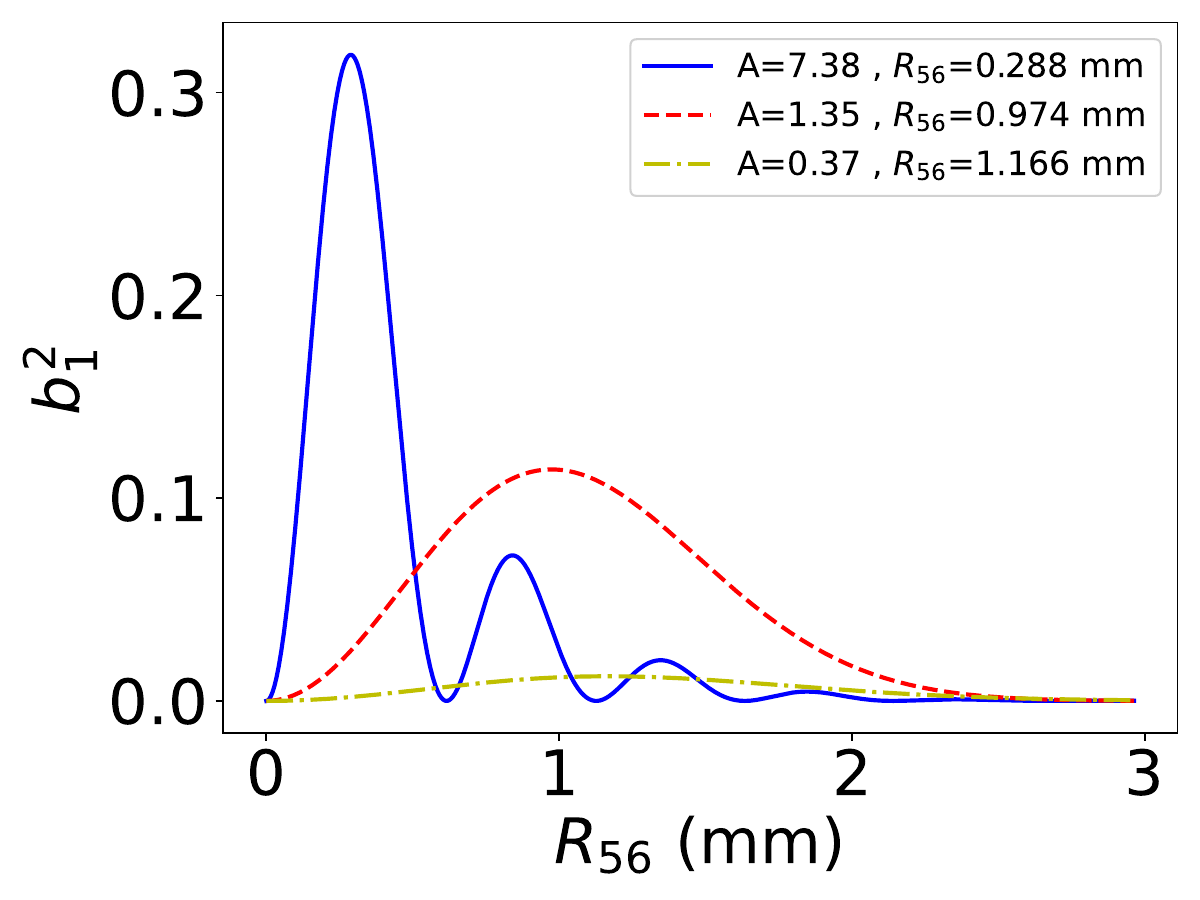}}
\caption{Simulated (a) and calculated (b) bunching factor squared as a function of dispersion strength under various phase shift values.}
\label{fig:9}
\end{figure}

The weak residual energy modulation amplitude is then indirectly inferred from the corresponding optimal $R_{56}$. Nevertheless, a theoretical lower limit, determined by the intrinsic relationship between $\Delta \gamma$ and $R_{56}$, still exists, as discussed in Sec.~\ref{sec:2}. The coherent undulator radiation method remains effective. The optimal $R_{56}$ obtained without a phase shifter is first taken as a reference, and the optimal $R_{56}$ values are subsequently scanned under different phase-shift conditions. 

For each setting of the phase shifter, the dispersive strength $R_{56}$ of the chicane is scanned, and the coherent radiation intensity downstream of the diagnostic undulator is recorded. As illustrated in Figs.~\ref{fig:8a} and \ref{fig:8b}, a coarse scan following the procedure described above allows the optimal demodulation phase to be identified. Taking three representative phase-shift settings, $\phi = 0\pi$, $0.82\pi$, and $0.96\pi$, as examples, the optimal dispersive strengths are found to be 0.31~mm, 0.98~mm, and 1.10~mm, respectively, corresponding to true energy modulation amplitudes of 7.23, 1.35, and 0.24.

To further improve the accuracy in evaluating weak residual energy modulation, time-dependent simulations are performed. In these simulations, the electron beam is modulated by a 50~MW seed laser with a pulse duration of 200~fs (FWHM). 
Figure~\ref{fig:9} shows the evolution of the fundamental bunching factor at the entrance of the diagnostic undulator as a function of $R_{56}$. 
The corresponding optimal $R_{56}$ values are 0.29~mm, 0.97~mm, and 1.17~mm, respectively. The observed differences in the optimal $R_{56}$ values are largely attributed to the fact that the seed laser pulse length is shorter than the electron bunch length \citep{Zeng2025}. Using the relationship between $\Delta\gamma$ and $R_{56}$ introduced earlier, the estimated energy modulation amplitudes are 7.38, 1.35, and 0.37, respectively.

It is evident that as the residual energy modulation amplitude decreases, its precise determination becomes increasingly sensitive to both the resolution of the $R_{56}$ scan and the accessible scanning range. In addition, the phase resolution of the phase shifter around odd multiples of $\pi$ directly affects the demodulation efficiency. To mitigate these effects, repeated scans can be performed in the vicinity of the optimal demodulation phase, and the above procedure can be iterated to reduce statistical uncertainties.

\begin{figure}
% \centering
\includegraphics[width=0.4\textwidth]{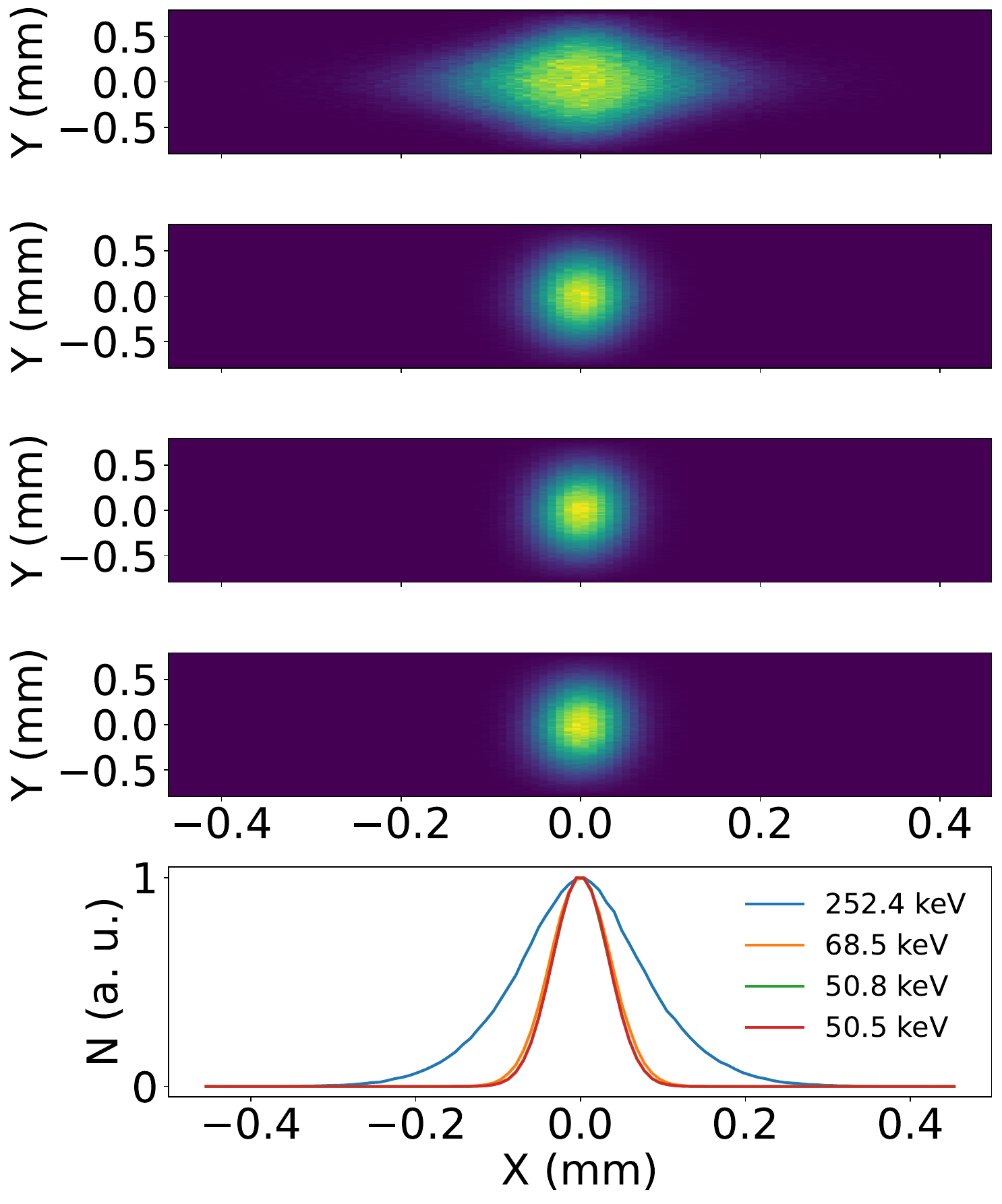}
\caption{Downstream XTDS images of the electron beam corresponding to various phase shifter settings. The final slice energy spread decreases progressively from top to bottom, leading to visibly distinct beam profiles.}
\label{fig:10}
\end{figure}

Two additional approaches are employed to indirectly evaluate the demodulation performance. First, the spectrum of the fundamental coherent radiation can be diagnosed using a downstream spectrometer. A weaker residual energy modulation leads to a pronounced reduction in the spectral intensity; for example, when the phase shift is varied from $0.82\pi$ to $0.96\pi$, the spectral intensity decreases by more than an order of magnitude. Moreover, in the weak-modulation regime, the residual energy modulation can be regarded as being induced by an effectively very weak seed laser, resulting in a significantly shortened radiation pulse duration and a correspondingly increased bandwidth. Second, the demodulated electron beam can be directly characterized using a transverse deflecting cavity combined with a dipole magnet downstream of the radiator, which enables slice energy spread measurements. By applying Eq.~(\ref{eq:3}), the residual energy modulation amplitude can be quantitatively estimated, and the detailed procedure is described in the next subsection. As shown in Fig.~\ref{fig:10}, as the phase shift is varied from $0\pi$ through $0.82\pi$ to $0.96\pi$, approaching the optimal demodulation value, the beam spot size on the downstream profile monitor screen gradually decreases from top to bottom, reflecting the progressive suppression of the energy modulation. The bottom image corresponds to a case obtained through a refined phase scan, for which the true residual energy modulation amplitude is 0.14, and the true slice energy spread is 50.3~keV.

\subsection{\label{sec:4d}Time-resolved measurements of slice energy spread}
The energy spread represents a critical parameter for accelerator commissioning and operation, typically measured using a TDS in combination with an energy-dispersive dipole magnet. This technique maps the electron beam's longitudinal phase space onto a diagnostic screen positioned within the dispersive section. The measured slice energy spread encompasses multiple contributions, including TDS-induced energy spread, beam betatron effects, screen resolution limitations, and the intrinsic energy spread, as expressed in \citep{Prat2020a,Tomin2021,Qian2022}:

\begin{equation}
\sigma_{\text{total}}^{2} = \sigma_{\text{scr}}^{2} + \frac{\varepsilon_{\mathrm{nl}} \beta_{\text{scr}}}{\gamma} + \left( D \frac{\sigma_{\gamma}}{\gamma} \right)^{2} + \left( D \frac{\sigma_{\gamma,\text{TDS}}}{\gamma} \right)^{2}
\end{equation}

\begin{figure}
% \centering
{
\includegraphics[width=0.5\textwidth]{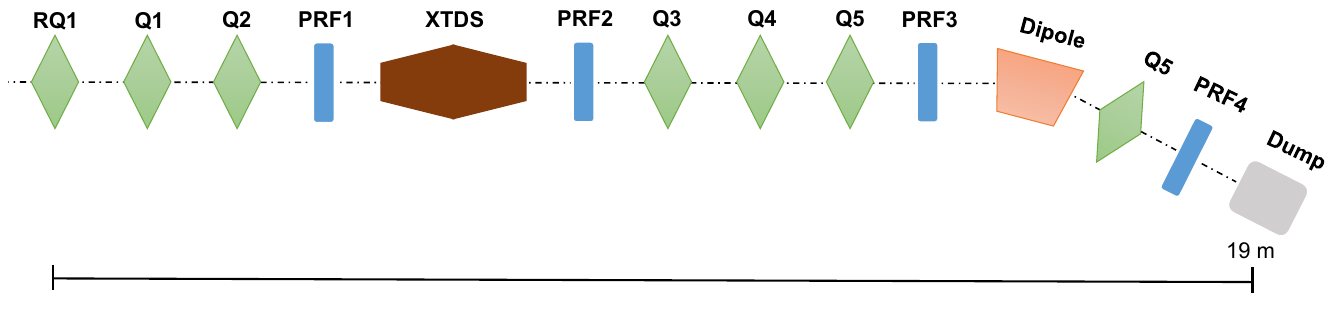}}
\caption{Schematic beamline for slice energy spread measurement at the SXFEL (element positions are not in proportion).}
\label{fig:11}
\end{figure}

\begin{figure}
\centering
\subfigure[\label{fig:12a}]{
\hspace*{-16pt}
\includegraphics[width=0.265\textwidth]{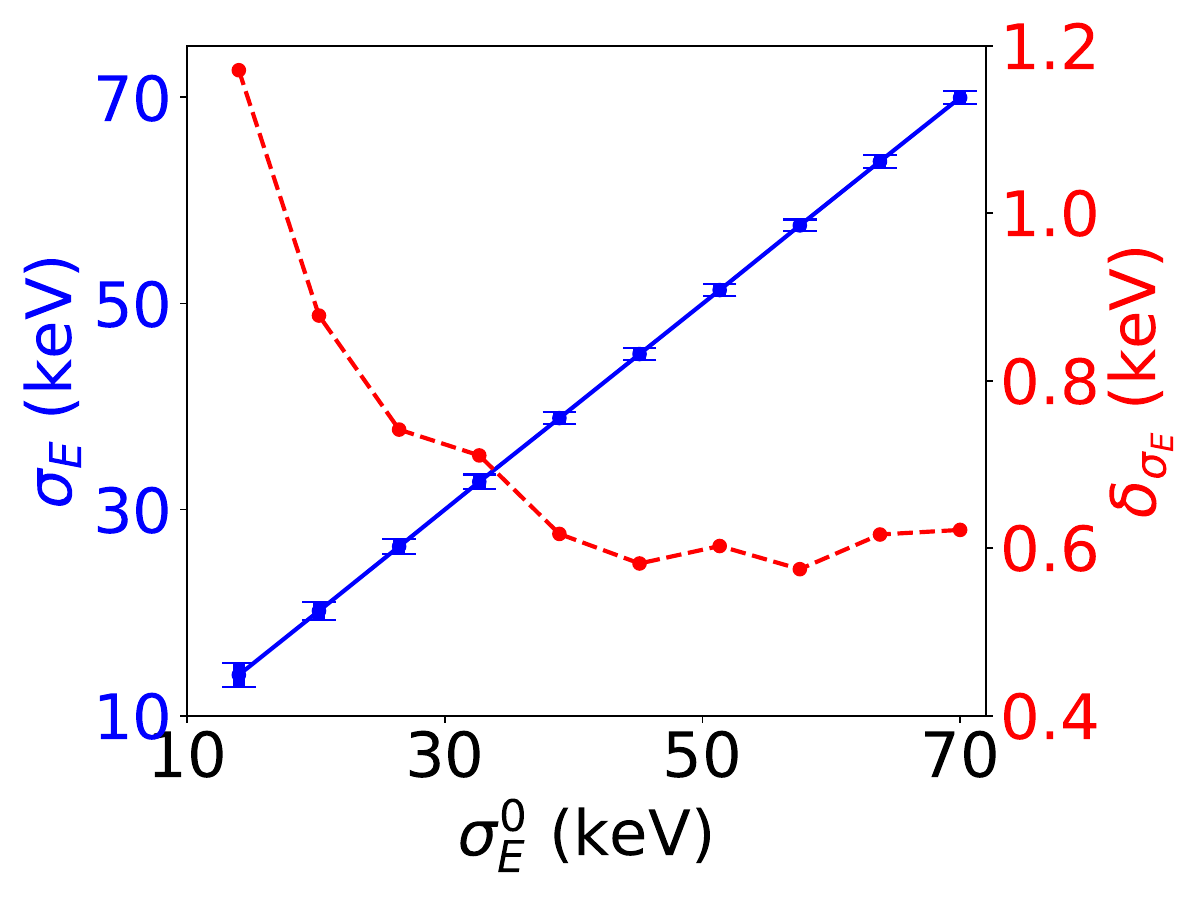}}
\subfigure[\label{fig:12b}]{
\includegraphics[width=0.225\textwidth]{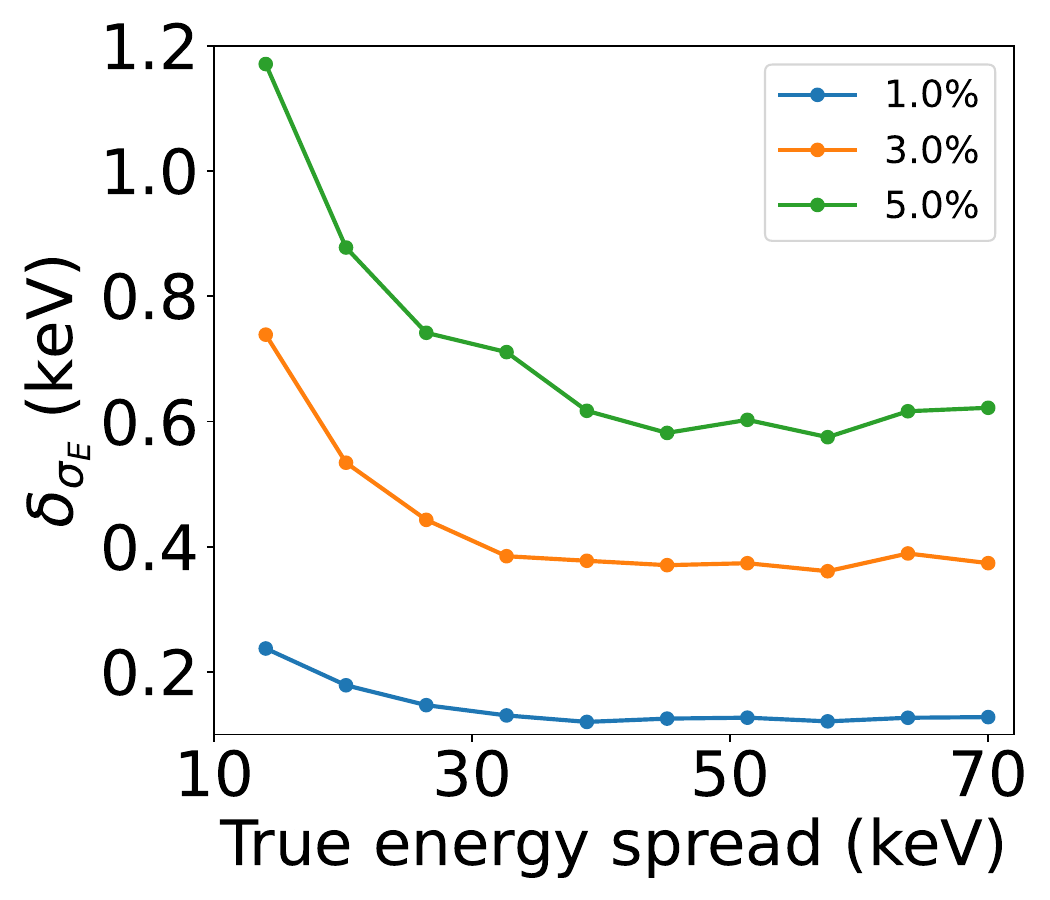}}
\caption{Simulations of the slice energy spread measurement: (a) deviation of the measured slice energy spread from the true value as a function of the absolute energy spread, assuming a 5\% RMS jitter in the beam spot size on the profile monitor screen; (b) dependence of the measurement accuracy on various RMS jitter levels of the beam spot size.}
\label{fig:12}
\end{figure}

Conventional methods for measuring slice energy spread involve scanning the TDS cavity voltage and varying beam parameters to separate the different contributions to the total measured energy spread on the screen, thereby extracting the intrinsic slice energy spread. The dispersion scan method implemented at the SXFEL seeding beamline achieves a measurement accuracy of 5\%@0.06\% (accuracy@relative energy spread). The schematic layout of the beamline at the radiator exit is shown in Fig.~\ref{fig:11}.

Prior to energy spread measurement, precise calibration of the quadrupoles, the X-band transverse deflecting structure (XTDS), and the analysis dipole along the diagnostic line is essential. Quadrupoles RQ1 and profile monitor screen PRF1 are employed to determine the beam's Twiss parameters. As is well established, the beam size observed in the dispersion direction at the profile monitor screen PRF4 results from a combination of the beam's intrinsic energy spread, its transverse quality (betatron), the XTDS kick, and the finite spatial resolution of the screen—30 $\mu$m in this case. The XTDS contribution can be isolated through voltage scanning, while the uncorrelated intrinsic energy spread is extracted via dispersion scans. The most challenging aspect involves decoupling the contributions from beam emittance and energy spread, which requires maintaining identical $\beta$-functions at PRF4 across different dispersion settings. In practice, this is accomplished by using quadrupoles Q4 and Q5 to match the beam optics, with only minimal adjustments to their strengths to reduce measurement errors.

Further error analysis demonstrates that the measurement uncertainty of the slice energy spread increases substantially as the absolute energy spread decreases, resulting in progressive deviation from the true value. As shown in Fig.~\ref{fig:12a}, when considering only a 5\% RMS jitter in beam spot size on PRF4, the measured energy spread values rapidly diverge from actual values below 30 keV. Specifically, at a nominal energy spread of 50 keV, a deviation of approximately 0.6 keV is observed. Figure~\ref{fig:12b} further indicates that under the current beam conditions, reducing the RMS jitter of the beam spot size on PRF4 yields measurements approaching the true energy spread value. Nevertheless, achieving such minimal spot size fluctuations under experimental conditions remains challenging, making it difficult to satisfy the ideal requirements for precise low-energy-spread measurements.
% Beam longitudinal phase space at the end of the radiator in various laser-induced energy modulation, with A = 7.22, 1.35, 0.38, 0.24, respectively.
\section{\label{sec:5}Demodulation undulator design}
\begin{figure}
{\includegraphics[width=0.25\textwidth]{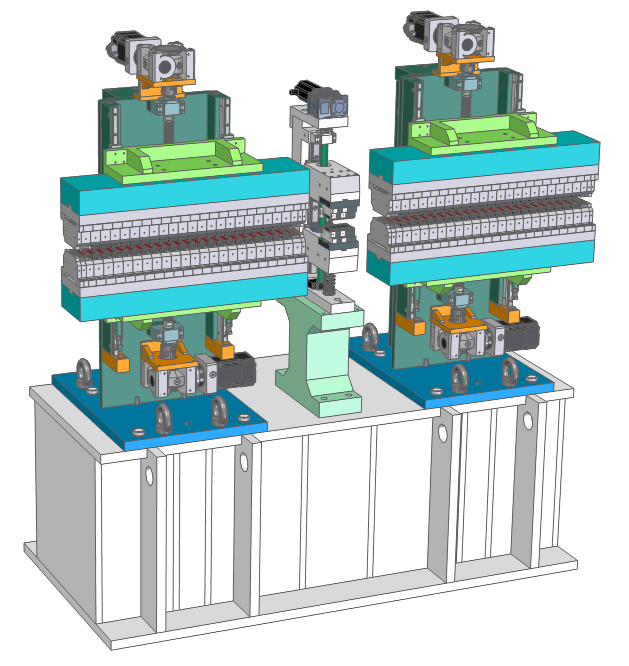}}
\caption{Mechanical layout of the proposed demodulation undulator, comprising two identical planar undulators separated by a tunable phase shifter.}
\label{fig:13}
\end{figure}

\begin{figure}
\subfigure[\label{fig:14b}]{
\includegraphics[width=0.4\textwidth]{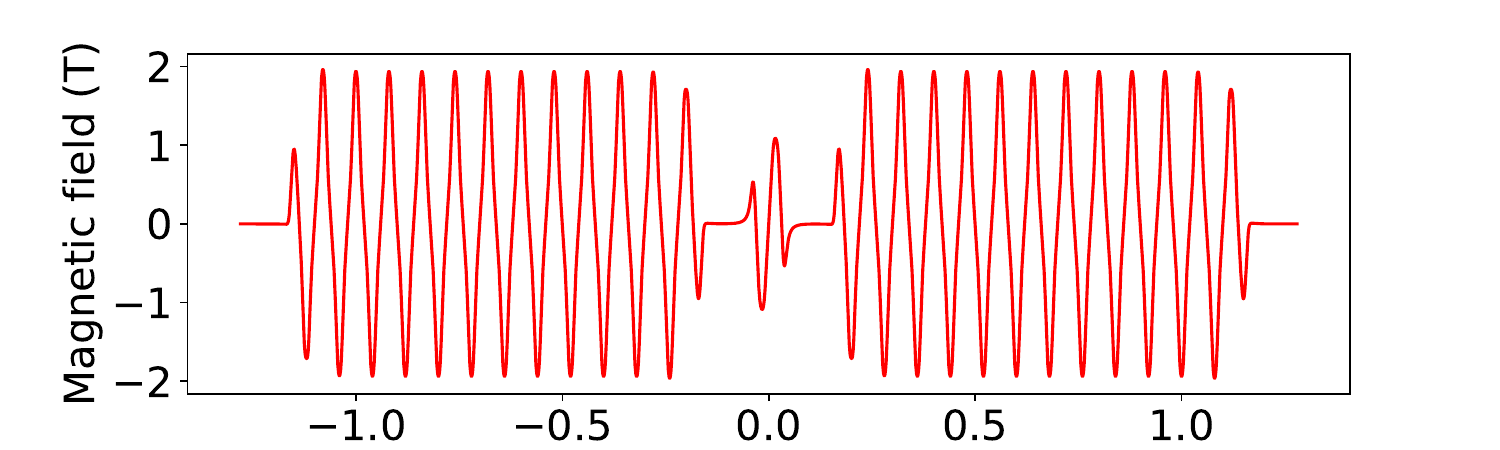}}
\subfigure[\label{fig:14c}]{
\includegraphics[width=0.4\textwidth]{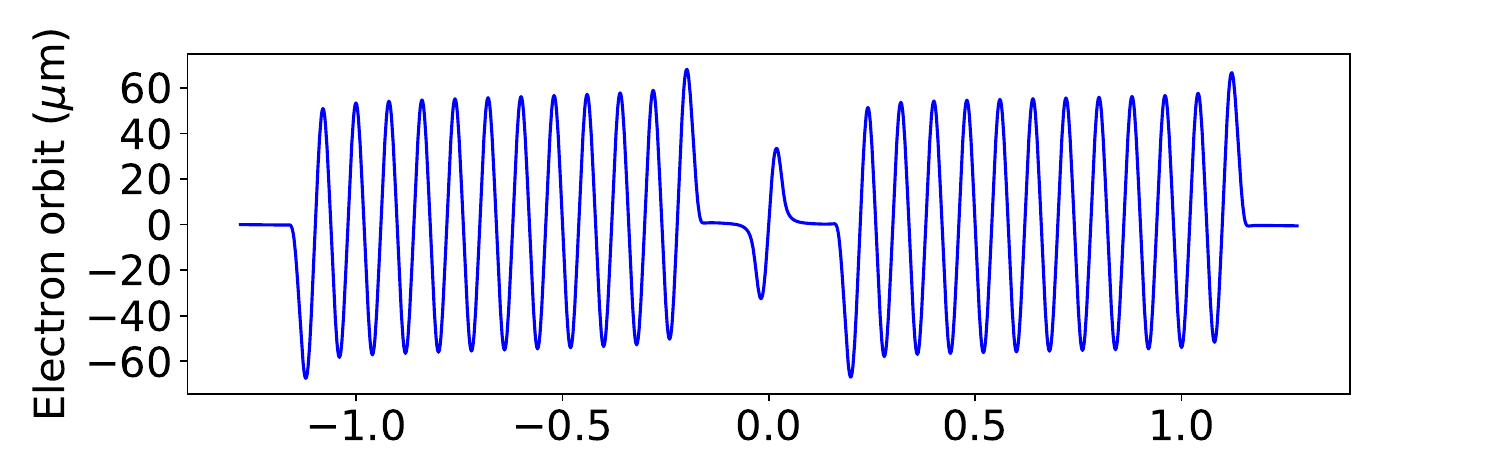}}
\caption{Calculated on-axis magnetic field and optimized electron trajectory through the demodulation undulator.}
\label{fig:14}
\end{figure}

To support future dedicated demodulation experiments, a compact demodulation undulator has been designed with an 80 mm period length, comprising two identical modulator segments (12 periods each) separated by a tunable phase shifter. The complete mechanical layout is shown in Fig.~\ref{fig:13}. Using the Opera electromagnetic simulation code \citep{opera_simulia}, we systematically optimized the undulator's effective magnetic field while minimizing demagnetization and fringe field effects. Figure~\ref{fig:14} presents the optimized on-axis magnetic field distribution and corresponding electron trajectories at a resonant radiation wavelength of 266 nm.

The demodulator's period count is intentionally limited to enable clear observation of the energy modulation-demodulation process and facilitate online validation of energy spread suppression. The phase shifter provides full $0$–$4\pi$ phase-shifting capability for a 1.4 GeV electron beam at 266 nm wavelength.

The demodulator employs permanent-magnet materials in an antisymmetric configuration, with identical period lengths and periods in both modulator segments. Careful optimization of the spacing between segments and their end-field distributions minimizes electromagnetic interference among components and reduces susceptibility to the geomagnetic field. This design maintains a stable beam orbit and transverse size across a wide operating range without requiring additional focusing elements between sections.

The U80 planar undulator utilizes an antisymmetric Halbach magnetic array of high-grade NdFeB permanent magnets and CoVFe soft magnetic poles, producing a near-ideal sinusoidal field profile with optimized strength, uniformity, and minimal phase error \citep{He2026}. Its mechanical structure consists of a high-stiffness base girder and C-shaped support frame, with symmetrically arranged ball screws and linear guides that suppress gap deformation from magnetic attraction forces and enable precise batch assembly \citep{Zhou2025}. Multi-axis servo systems and linear absolute encoders provide precise control of gap adjustment, taper control, and center shifting with micrometer-level accuracy and repeatability \citep{Lei2024}.

The PS80 phase shifter features a compact magnetic structure with a high-precision linear motion mechanism, introducing controlled phase delays through minimal mechanical displacements. Its control interface is fully compatible with the U80, enabling seamless integration into a unified operational framework.

The complete system integrates two identical U80 undulators in series with the PS80 phase shifter positioned centrally between them. All three components share a common mechanical support, vacuum chamber, and alignment reference, forming a compact, functionally integrated unit. This configuration allows the electron beam to traverse the front U80, PS80, and rear U80 sequentially, with the U80 segments providing primary radiation gain while PS80 ensures optimal phase matching for enhanced output radiation intensity, coherence, and spectral properties. The three-segment architecture exhibits high modularity in mechanical design, magnetic configuration, and control logic, enabling repeated deployment along the beamline with consistent performance, unified control, and simplified maintenance.

% \section{\label{sec:6}Experimental results}

\section{\label{sec:6}Preliminary experiments at SXFEL}
\begin{figure*}
\subfigure[\label{fig:exp_res_a}]
{\includegraphics[width=0.32\textwidth]{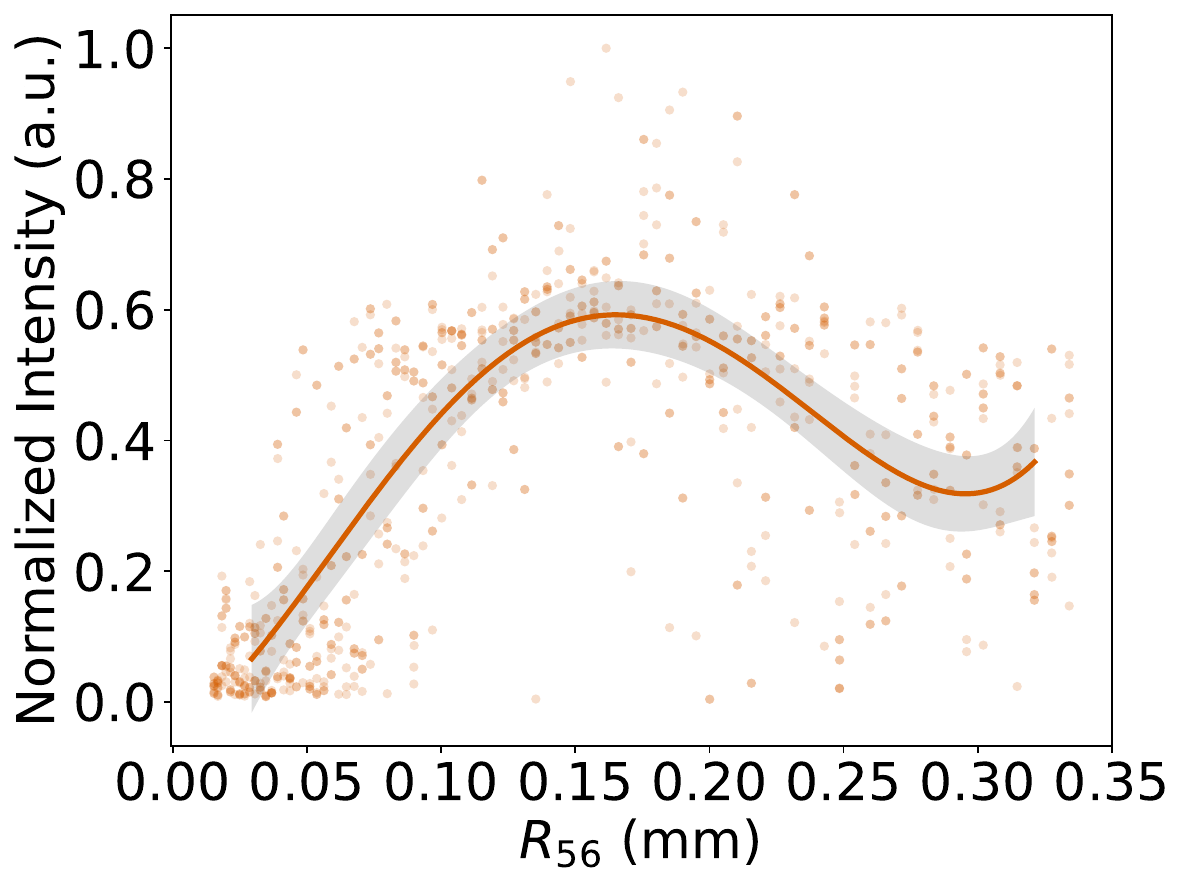}}
\subfigure[\label{fig:exp_res_b}]
{\includegraphics[width=0.32\textwidth]{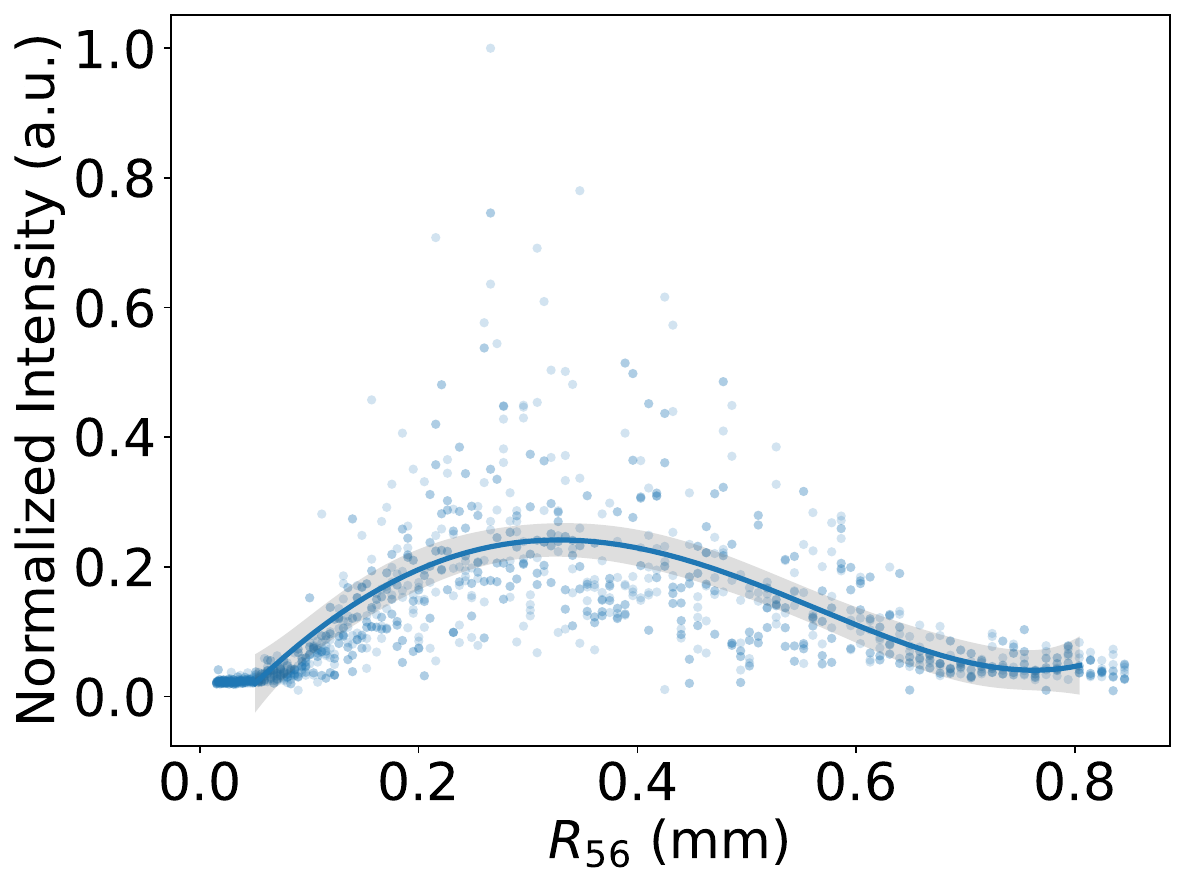}}
\subfigure[\label{fig:exp_res_c}]
{\includegraphics[width=0.32\textwidth]{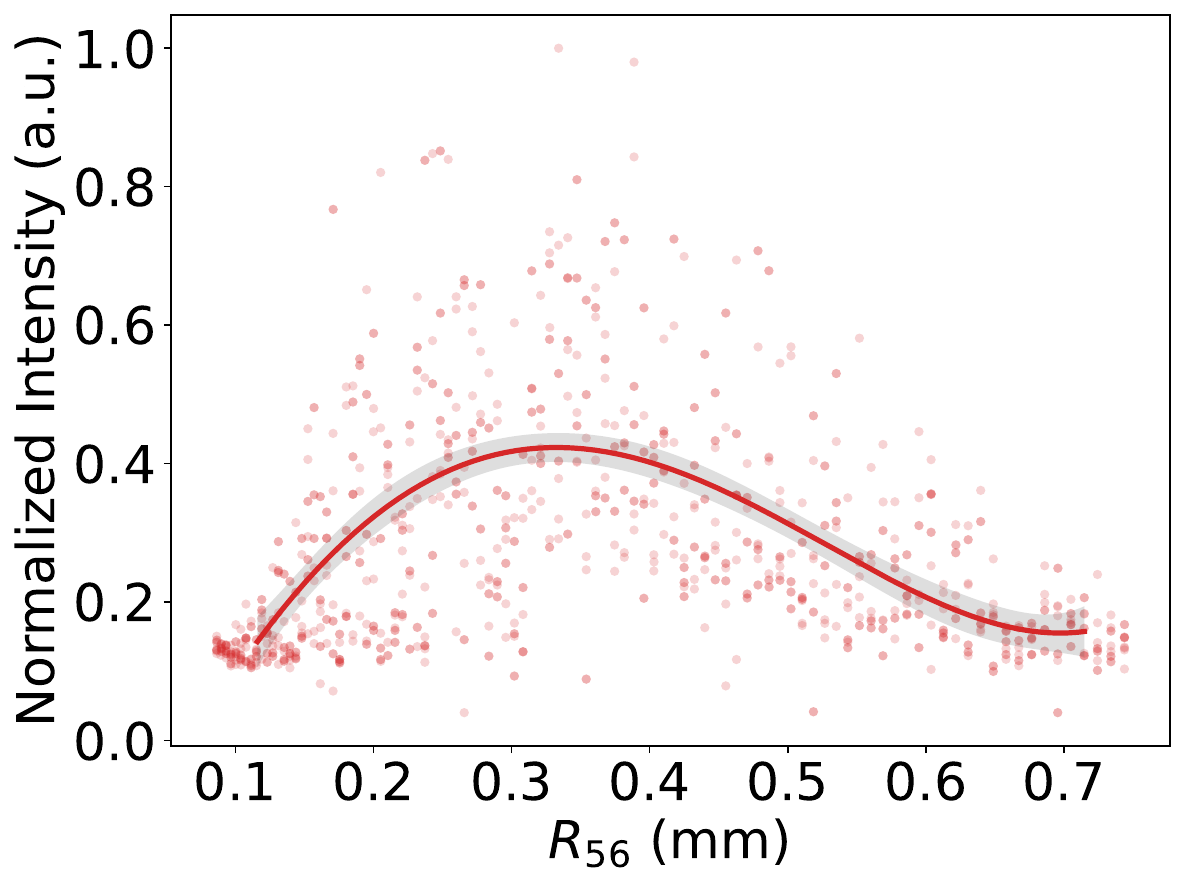}}
\caption{Estimation of residual energy modulation from coherent radiation measurements. 
The coherent radiation signal is shown as a function of the dispersion strength for 
(a) the zero-phase condition, 
(b) the demodulation-phase condition, and 
(c) the single-modulator reference case obtained by detuning the second modulator. 
The red curves represent the fitted profiles, and the gray shaded regions indicate the 95\% confidence intervals. 
The optimal dispersion strengths extracted from the fits are $R_{56}=0.165$~mm, $0.328$~mm, and $0.333$~mm for panels (a), (b), and (c), respectively. 
The larger optimal $R_{56}$ values in (b) and (c) indicate reduced energy modulation amplitudes compared with the zero-phase condition in (a).}
\label{fig:exp_res}
\end{figure*}

Preliminary experimental studies of energy modulation suppression were carried out at SXFEL using the existing seeding beamline. These measurements were not performed with the dedicated U80 demodulation undulator discussed in Sec.~\ref{sec:5}; instead, we used the latest available SXFEL layout, whose basic configuration is similar to the modulation--demodulation scheme considered in this work. The electron beam energy was 890~MeV and the peak current was around 500~A. Two 68~mm-period undulators, each with an effective magnetic length of about 4~m, were used as the two modulation sections, with a phase shifter installed between them. A 50~mm-period undulator was used as the diagnostic undulator.

First, the coherent undulator radiation method was applied to determine the slice energy spread and modulation amplitude. With both modulation sections tuned to the seed-laser resonance and the phase shifter optimized, the electron beam was modulated by an ultraviolet seed laser at 266~nm with a pulse duration of approximately 200~fs. Coherent-radiation measurements were then performed at two seed-laser pulse energies, 39.1~$\mu$J and 15.1~$\mu$J. The diagnostic undulator was tuned to the second harmonic of the seed laser, and the dispersive strength was scanned to measure the coherent radiation intensity. The measured coherent radiation intensity and the fitted curves exhibit distinct optimal dispersion strengths for the two seed-laser energies. By combining the optimal dispersion values with the analytical relation between the slice energy spread and the modulation amplitude, the slice energy spread was estimated to be approximately 87~keV.

After establishing the slice-energy-spread and modulation-amplitude reference, the phase dependence of the modulation suppression was examined by scanning the phase shifter between the two modulation sections. The seed-laser pulse energy was fixed at 15.1~$\mu$J, and both modulation sections participated in the laser--beam interaction. The coherent radiation signal from the diagnostic undulator was recorded while scanning the gap of the phase shifter between the two modulation sections. To search for the demodulation condition with sufficient signal-to-noise ratio, the dispersive strength was set to 0.086~mm, slightly away from the theoretical optimum. A phase-shifter gap of 24~mm corresponds to the zero-phase condition, where the electron beam undergoes standard energy modulation. When the gap was increased to about 33~mm, corresponding to a demodulation phase close to $0.9\pi$, the coherent radiation intensity was clearly reduced. This reduction indicates that the induced energy modulation was suppressed relative to its maximum value, consistent with the expected modulation--demodulation process.

The residual modulation amplitude was further estimated from coherent radiation measurements using an 80~mm-period diagnostic undulator operated at the fundamental wavelength. As shown in Fig.~\ref{fig:exp_res}, the coherent radiation signal was measured as a function of the dispersion strength under three representative conditions. For the zero-phase condition, shown in Fig.~\ref{fig:exp_res_a}, the optimal dispersion strength was approximately 0.165~mm. When the phase shifter was tuned close to the demodulation phase, as shown in Fig.~\ref{fig:exp_res_b}, the optimal dispersion strength increased to approximately 0.328~mm, indicating a reduced residual modulation amplitude. To provide a reference corresponding to a single modulation section, the second 68~mm-period undulator was detuned so that only the first modulation section contributed to the laser--beam interaction. Although the coherent radiation intensity was significantly reduced in this case, the optimal dispersion strength could still be identified as approximately 0.333~mm, as shown in Fig.~\ref{fig:exp_res_c}, which is close to that obtained under the demodulation-phase condition.

Using the analytical relation introduced in Sec.~\ref{sec:3}, the estimated modulation amplitudes for the zero-phase condition, the demodulation-phase condition, and the single-modulator reference case are approximately 4.1, 1.3, and 1.3, respectively. These results show that the energy modulation amplitude was reduced by about a factor of three relative to its maximum value and recovered close to the single-modulator level. Although the present experiment was performed with the existing SXFEL configuration rather than a dedicated demodulator optimized for complete cancellation, the observation provides experimental evidence of energy modulation suppression in a seeded FEL and supports systematic demodulation studies on a dedicated platform.

For weak residual modulation, the measurement accuracy is sensitive to the $R_{56}$ scan resolution, the accessible scan range, and shot-to-shot fluctuations. Repeated scans around the optimal demodulation phase, combined with statistical fitting, can reduce the resulting uncertainty. Two complementary diagnostics provide additional cross-checks: spectral measurements of the coherent fundamental radiation downstream of the diagnostic undulator, and time-resolved slice-energy-spread measurements. Combining these measurements enables quantitative evaluation of the demodulation efficiency and the associated suppression of the slice energy spread.
% For weak residual modulation, the measurement accuracy is sensitive to the $R_{56}$ scan resolution, the accessible scan range, and shot-to-shot fluctuations. Further measurements will therefore use repeated scans around the optimal demodulation phase and statistical fitting to reduce uncertainty. Two complementary diagnostics will also be used for cross-checking: spectral measurements of the coherent fundamental radiation downstream of the diagnostic undulator, and time-resolved slice-energy-spread measurements with a transverse deflecting cavity and a dipole spectrometer. Combining these measurements will allow a quantitative evaluation of the demodulation efficiency and the associated suppression of the slice energy spread.

\section{\label{sec:7}Conclusions and perspectives}
We have investigated energy modulation and demodulation in seeded FELs using a demodulation undulator configuration consisting of two modulators separated by a tunable phase shifter. Analytical analysis shows that a $\pi$ phase delay enables the laser-beam interaction to nearly reverse the initial modulation, thereby suppressing the residual energy modulation. Three-dimensional simulations based on representative SXFEL parameters confirm that substantial suppression can be achieved under optimized conditions, reaching the weak-modulation regime in which accurate characterization becomes experimentally demanding.

To support quantitative characterization of this process, we studied complementary diagnostics based on coherent undulator radiation and dispersion-scan measurements. These methods provide both indirect and direct routes for evaluating the residual energy modulation and the associated energy-spread growth. A compact demodulation undulator was also designed as a future dedicated platform for controlled demodulation experiments.

Preliminary experiments performed at SXFEL using the existing seeding beamline reveal clear signatures of energy modulation suppression. Under the available experimental configuration, the modulation amplitude was reduced by approximately a factor of three and recovered close to the single-modulator level. Although this configuration was not optimized for complete demodulation, the results provide experimental evidence of energy-modulation suppression associated with the modulation-to-demodulation process in a seeded FEL and validate the main diagnostic workflow.

Precise control of laser-induced energy modulation will become increasingly important as seeded FEL facilities advance toward high repetition rates and short wavelengths, where stringent constraints are imposed on the induced energy spread in the modulator \citep{Yan2021,qi2025}. The framework developed here provides a basis for systematic demodulation studies at SXFEL and can be adapted to other seeded FEL facilities employing similar laser-beam manipulation and diagnostic schemes, such as FERMI \citep{Allaria2025} and DCLS \citep{Li2025}.

\begin{acknowledgments}
The authors thank Zheng Qi, Zhen Wang, Chao Feng, Chunlei Li, Wenyi Yin, Xingtao Wang, Guanglei Wang, Duan Gu, and Si Chen for their valuable discussions and interest in this work. The authors are particularly grateful to Jiawei Yan for his valuable discussions, insightful suggestions, and continued support throughout this work. This work was supported by the National Key Research and Development Program of China (2024YFA1612101, 2024YFA1612104), the National Natural Science Foundation of China (12125508, 12541503, 12475322, 12205358), the Shanghai Pilot Program for Basic Research – Chinese Academy of Sciences, Shanghai Branch (JCYJ-SHFY-2021-010), the Shanghai Natural Science Foundation (23ZR1471300), the Shanghai Municipal Science and Technology Major Project, the Innovation Program of Shanghai Advanced Research Institute, CAS (2025CP006), and the China Postdoctoral Science Foundation (2025M770914).
\end{acknowledgments}

% \nocite{*}

\bibliography{reference}% Produces the bibliography via BibTeX.

\end{document}